\documentclass[sigplan,10pt]{acmart}
\renewcommand\footnotetextcopyrightpermission[1]{}
\setcopyright{none}
\acmDOI{}
\acmISBN{}
\acmConference[]{}{}{}
\acmBooktitle{}
\makeatletter
\renewcommand\@mkbibcitation{}
\def\@mkauthors{%
  \begingroup
  \hsize=\textwidth
  \global\setbox\mktitle@bx=\vbox{%
    \noindent\unvbox\mktitle@bx
    \par\medskip
    \begin{center}
      {\Large
      Akhmed Sakip, Erland Hilman Fuadi, Omar Sayedelahl, Zonghang Li,\\
      Jianshu She, Alham Fikri Aji, Steve Liu, Eric Xing, Qirong Ho\par}
      \vspace{0.4em}
      {\normalsize Mohamed bin Zayed University of Artificial Intelligence\par}
    \end{center}
    \par\medskip}%
  \endgroup}
\makeatother

\usepackage{booktabs}
\usepackage{tikz}
\usepackage{subcaption}
\usepackage{algorithm}
\usepackage{algpseudocode}
\usepackage{amsmath}

\newcommand{\Bg}{B_g}
\newcommand{\Bm}{B_m}
\newcommand{\Bcrit}{B_{\mathrm{crit}}}
\newcommand{\SE}{\mathrm{SE}}
\newcommand{\GNS}{\phi}
\newcommand{\sys}{\textsc{Copus}}

\begin{document}

\title{COPUS: Co-adaptive Parallelism and Batch Size Selection in Large Language Model Training}

\author[Sakip et al.]{Akhmed Sakip}
\author{Erland Hilman Fuadi}
\author{Omar Sayedelahl}
\author{Zonghang Li}
\author{Jianshu She}
\author{Alham Fikri Aji}
\author{Steve Liu}
\author{Eric Xing}
\author{Qirong Ho}
\authorsaddresses{}

\begin{abstract}
Training large language models requires jointly configuring two
interdependent aspects of the system: the global batch size, which
governs statistical efficiency, and the 3D parallelism strategy (data,
tensor, pipeline), which governs hardware throughput. Existing approaches
make these decisions independently: optimization work adapts the batch
size to track the evolving critical batch size while keeping
parallelism fixed, and systems work selects the fastest parallelism
for a given fixed batch size without anticipating that the optimal batch size could change. We show that these decisions are tightly coupled: the
throughput-optimal parallelism strategy may shift as the global batch size
changes, so any method that fixes one while adapting the other operates
with a suboptimal configuration for part of the training run.

We present \sys{}, the first system that adaptively tunes not only the
global batch size but also the throughput parameters, parallelism
strategy and micro-batch size, as training evolves. \sys{} is guided
by \emph{Goodput}, the product of throughput and statistical
efficiency, which models both hardware and statistical effects jointly
and directly measures useful convergence per unit of wall-clock time.
Unlike prior adaptive batching approaches that maximize per-sample
efficiency alone, \sys{} co-optimizes all three parameters to maximize
the rate at which training converges. The system combines online gradient noise scale estimation under 3D
parallelism with throughput-aware evaluation of candidate
configurations to continuously select the best one, and supports
efficient reconfiguration of both batch size and parallelism during
training. We evaluate \sys{} on LLM pre-training workloads
across 1--4 nodes of 8$\times$H100 and 8$\times$MI210 GPUs and model
sizes from 3B to 32B parameters, demonstrating average
time-to-convergence speedups of \textbf{3.9--8.0\%} over the fastest
baseline across four configurations, with peak gains up to
\textbf{11.1\%}, including system overheads.
\end{abstract}

\maketitle
\begingroup
\renewcommand{\thefootnote}{}
\footnotetext{Correspondence: Akhmed Sakip (\href{mailto:akhmed.sakip@mbzuai.ac.ae}{akhmed.sakip@mbzuai.ac.ae}).}
\addtocounter{footnote}{-1}
\endgroup
\pagestyle{plain}

\begin{figure}[t]
    \centering
    \small
    \includegraphics[width=\linewidth]{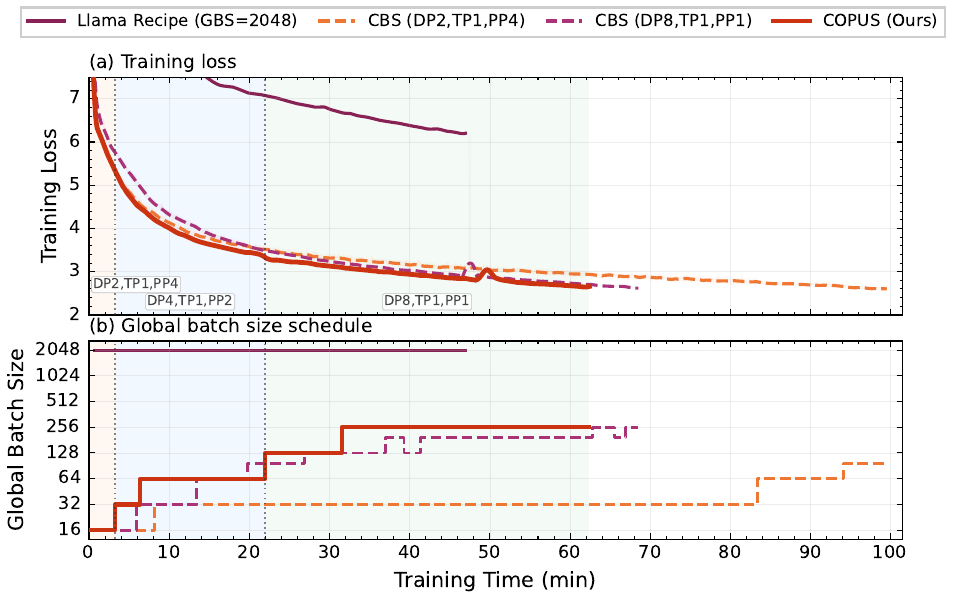}
    \caption{\sys{} adaptation trajectory on 3B / 1$\times$8 H100. (a)~Training loss. (b)~Batch size schedule. Colored regions and dotted lines show \sys{}'s three parallelism strategies. CBS baselines adapt batch size but keep parallelism fixed.}
    \label{fig:behavior_trajectory}
\end{figure}

\section{Introduction}
\label{sec:introduction}

The training of large language models (LLMs) is among the most computationally intensive workloads in modern computing~\cite{llama3,gemini}, with runs lasting weeks on clusters of thousands of accelerators. Two fundamental decisions govern the efficiency of such runs: the \emph{global batch size} ($\Bg$) and its decomposition into \emph{micro-batches} ($\Bm$), which control how much data is consumed per optimization step, and the \emph{3D parallelism strategy} $S = (d, t, p)$ for data, tensor, and pipeline parallelism, which determines how the model and data are distributed across hardware. These decisions are deeply interdependent, yet current training practices treat them in isolation. In state-of-the-art distributed training frameworks like Megatron-LM~\cite{megatron-lm} and DeepSpeed~\cite{deepspeed}, the 3D parallelism topology must be statically declared at initialization.

Consequently, even the most advanced industrial pre-training runs often rely on compromises that can become suboptimal. For instance, during the training of AI2's OLMo 65B~\cite{groeneveld2024olmoacceleratingsciencelanguage}, the training recipe relies on a static configuration where the batch size is scheduled to repeatedly double (e.g., from roughly 2M to roughly 16M tokens) to track the evolving critical batch size. However, because the framework's parallel mesh is locked at launch, the system can only increase gradient accumulation steps to absorb the larger global batch size, leaving the underlying 3D parallelism strategy fixed. Similarly, Meta's LLaMA-3 405B~\cite{llama3} varied its batch size over training, but the reported training recipe treats batch-size scheduling and parallelism planning as separate engineering choices rather than as one co-adaptation problem. This mismatch can leave hardware throughput on the table: for a given model-hardware pair, when the adaptive batch-size trajectory crosses regimes where different parallelism strategies are fastest, a fixed layout cannot follow the hardware-optimal configuration.
On the optimization side, prior work on adaptive batch sizing~\cite{gns, cbs_revisited, adaptdl} shows that the statistically efficient batch size is not fixed: it typically grows during training. On the systems side, parallelism optimizers such as Alpa~\cite{alpa} and Galvatron~\cite{galvatron} search for the fastest execution strategy for a given fixed batch size. These two lines of work solve complementary but incomplete problems. If $\Bg$ changes during training, the throughput-optimal parallel strategy may change accordingly.

The core observation of this work is that these two decisions are \emph{coupled}: the throughput-optimal parallelism strategy depends strongly on the current batch size. \autoref{fig:behavior_trajectory} previews how \sys{} responds on a 3B model trained on 8 H100 GPUs: as the batch size grows during training, the system transitions through three parallelism strategies, each matched to the current batch size regime. Static baselines that lock parallelism at initialization cannot follow these shifts and lose throughput as the batch size outgrows their initial configuration. To illustrate the scale of the mismatch, \autoref{fig:behavior_trajectory} also includes a Llama-style training recipe that fixes the batch size at 2048 (${\sim}$4M tokens) with a tuned peak learning rate of $3{\times}10^{-4}$ and a ${\sim}$41M-token warmup. This fixed large-batch configuration reaches a loss of 6.2 after 47 minutes of training; \sys{} reaches the same loss in under 2 minutes, a roughly 30$\times$ gap that underscores how much wall-clock time large fixed batch sizes waste during early pre-training.

\paragraph{The right metric: Goodput.} Existing approaches optimize incomplete objectives. Adaptive batch sizing methods~\cite{gns,cbs,cbs_revisited,cabs} maximize \emph{statistical efficiency} ($\SE$, convergence per sample processed) but ignore processing time on real hardware. Parallelism optimizers~\cite{alpa,galvatron,flexflow,unity} maximize \emph{system throughput} ($T$, samples per second) but ignore the statistical properties of the batch size. What practitioners actually care about is \emph{convergence per unit time}, which is the product of these two quantities. Pollux~\cite{adaptdl} introduced the \emph{Goodput} metric to capture this in cluster scheduling, where it was used to co-adaptively tune batch size and scale resources across competing jobs. We argue that Goodput is better suited for batch size tuning in LLM training than CBS-only approaches, because it accounts for both the statistical and hardware sides of candidate configurations:
\begin{equation}
  \mathrm{Goodput}(S, \Bg, \Bm, H) = \mathrm{T}(S, \Bg, \Bm, H) \times \SE(\Bg).
  \label{eq:goodput_intro}
\end{equation}
By maximizing Goodput, \sys{} makes the batch size and parallelism decisions jointly: the throughput term enters the batch size decision itself, rather than being considered only after $\Bg$ has already been chosen. We further extend Pollux's formulation with a learning rate correction for Adam's square-root LR scaling \cite{sqrt_lr}, which CBS-only approaches cannot cleanly incorporate (\S\ref{sec:goodput}).

A practical concern with GNS-based approaches is that the raw GNS value requires a scaling factor to translate into a batch size, and this factor is not known a priori~\cite{gns,cbs_revisited,layernorm}. We argue that, given the same default scaling factor, Goodput-based selection produces better results than directly setting $\Bg$ from the GNS estimate. The key reason is that the throughput surface provides an independent constraint: even when the statistical efficiency curve is shifted due to scaling factor uncertainty, the throughput component partially counteracts this, reducing how far the selected batch size can deviate (\S\ref{sec:goodput}).

\paragraph{The \sys{} system.}
We present \sys{}, the first system to dynamically co-optimize $\Bg$, $\Bm$, and $S$ during a single LLM training run. \sys{} treats the training configuration as a dynamic tuple $(S, \Bg, \Bm)$ that evolves to maximize Goodput during training. The system consists of three components:

\begin{enumerate}
  \item A \textbf{3D-parallel-aware GNS estimator} integrated into the Megatron-LM training loop. Unlike prior estimators that assume pure data parallelism~\cite{adaptdl}, ours correctly handles gradient accumulation (temporal variance across micro-batches) and 3D parallelism (spatial variance across DP ranks), with near-zero overhead.

  \item A \textbf{Goodput orchestrator} that periodically combines the
  online GNS estimate with a pre-measured throughput lookup table to
  evaluate Goodput across all candidate $(S, \Bg, \Bm)$ configurations.
  It selects the configuration with maximum Goodput and performs the
  appropriate reconfiguration operation.

  \item An \textbf{adaptive training core} that supports both in-process
  batch size changes and parallelism strategy changes via online state
  resharding. For parallelism changes, this reduces reconfiguration
  latency by $2$--$16\times$ compared to full checkpoint-restart, the
  conventional way to change the parallelism strategy in standard
  training stacks that cannot reshard training state in place.
\end{enumerate}

\paragraph{Contributions.} We make the following contributions:
\begin{itemize}
  \item \textbf{The batch-parallelism coupling.} We empirically
  demonstrate that the throughput-optimal 3D parallel strategy may
  depend strongly on the global batch size
  (\autoref{fig:throughput_coupling}), and that this coupling causes
  methods optimizing batch size or parallelism in isolation to operate
  with a suboptimal configuration as training progresses
  (\S\ref{sec:motivation}).

  \item \textbf{Goodput-driven co-optimization.} We adopt the Goodput
  metric from Pollux~\cite{adaptdl} and formulate the joint selection
  of $(\Bg, \Bm, S)$ as a Goodput maximization problem. We show that
  Goodput-based selection outperforms CBS-only batch size selection
  because it accounts for both the statistical and hardware properties
  of candidate configurations (\S\ref{sec:goodput}).

  \item \textbf{The \sys{} system.} We design and implement \sys{},
  which integrates a 3D-parallel-aware GNS estimator, a Goodput
  orchestrator, and an adaptive training core with in-process
  reconfiguration support. The system is implemented as a fork of
  Megatron-LM (\S\ref{sec:design}, \S\ref{sec:implementation}).
\end{itemize}

We evaluate \sys{} on LLM pre-training across 1--4 nodes of
8$\times$H100 GPUs and model sizes from 3B to 32B parameters. We
compare against baselines where throughput configurations are fixed
throughout training (static parallelism with adaptive or fixed batch
size) and baselines where the batch size is adapted via CBS-only
methods that ignore throughput. \sys{} reduces time-to-convergence by
\textbf{3.9--8.0\%} on average compared to these baselines, with
peak gains up to \textbf{11.1\%}, including system overheads.

\section{Background and Motivation}
\label{sec:motivation}

Reducing time to convergence in LLM training requires reasoning about two coupled questions: how the job is executed on hardware, and how the chosen batch size affects optimization progress. The first is governed by the 3D parallelism strategy and micro-batch decomposition; the second by the global batch size. This section reviews these two sides and connects them through the key observation of this paper: the throughput-optimal parallel strategy changes as the statistically efficient batch size changes.

\subsection{LLM Training with 3D Parallelism}
\label{subsec:3d_parallel}

We denote a 3D parallel execution strategy by $S=(d,t,p)$, where $d$, $t$, and $p$ are the degrees of data, tensor, and pipeline parallelism, with $d \times t \times p = N_{\text{GPUs}}$. For a fixed global batch size $\Bg$, training also chooses a micro-batch size $\Bm$, and the number of gradient-accumulation steps is
\[
\mathrm{GA} = \frac{\Bg}{d \cdot \Bm}.
\]
This decomposition separates optimization-side and execution-side effects. The global batch size $\Bg$ primarily governs the statistical behavior of training, while $S$ and $\Bm$ determine memory footprint, communication pattern, pipeline efficiency, and device utilization. Data parallelism is typically most efficient when each replica has enough local work; tensor parallelism trades additional intra-layer communication for lower per-device memory; and pipeline parallelism trades stage-level concurrency against pipeline bubbles, making $\Bm$ and gradient accumulation important throughput knobs. Thus, the DP-dominant strategy is not always feasible: large models often require minimum tensor or pipeline parallelism degrees to fit the model, activation, and optimizer states in GPU memory, so \sys{} searches only among memory-feasible 3D strategies. Existing systems such as Megatron-LM~\cite{megatron-lm} and DeepSpeed~\cite{deepspeed} expose these choices as execution parameters that are usually selected before training and then kept fixed.

\begin{figure}[t]
    \centering
    \includegraphics[width=\columnwidth]{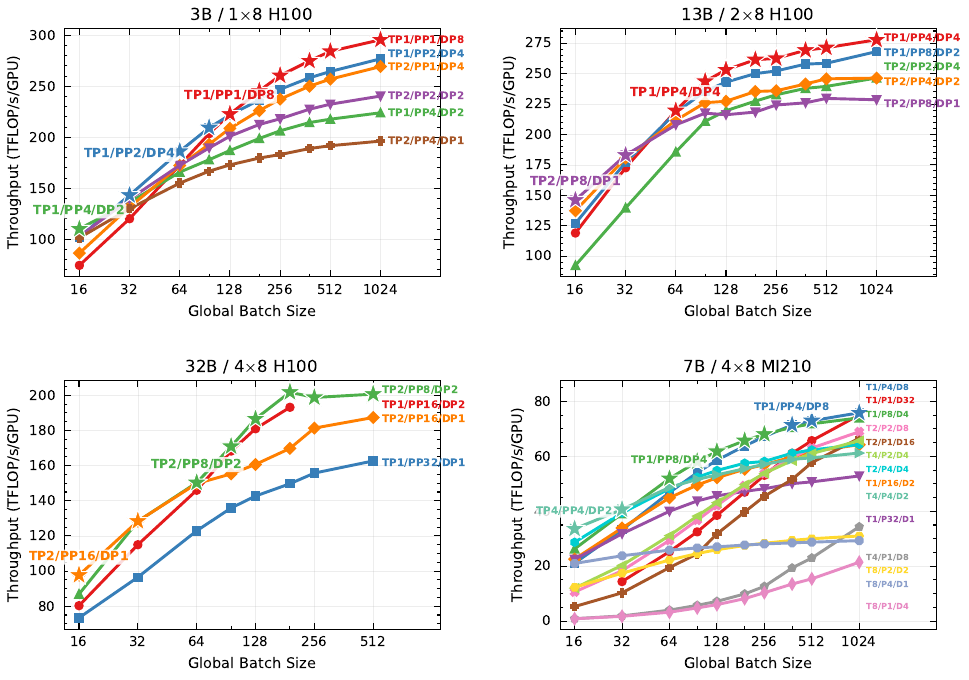}
    \caption{Throughput as a function of global batch size ($\Bg$, log scale) for all evaluated 3D parallel strategies across four hardware configurations. Each curve is a distinct $(t,p,d)$ configuration; star markers indicate the throughput-optimal strategy at each $\Bg$.}
    \label{fig:throughput_coupling}
\end{figure}

\subsection{Statistical Efficiency and Critical Batch Size}
\label{subsec:se_cbs}

While $S$ and $\Bm$ determine how efficiently the hardware executes a step, the global batch size $\Bg$ determines how much useful optimization progress that step makes. Following Pollux~\cite{adaptdl}, we use \emph{statistical efficiency} to denote convergence progress per sample processed. Small batches are often more sample-efficient, while the benefit of larger batches eventually saturates.

The standard online proxy for this behavior is the \emph{Gradient Noise
Scale} (GNS)~\cite{gns}, defined as:
\begin{equation}
  \phi = \frac{\mathrm{tr}(\Sigma)}{\lVert G \rVert^2},
  \label{eq:gns_background}
\end{equation}
where $G$ is the true gradient and $\Sigma$ is the gradient covariance.
Intuitively, larger $\phi$ means a lower gradient signal-to-noise
ratio, so averaging more samples remains useful over a larger
batch-size range. It therefore provides a proxy for the \emph{critical
batch size} $\Bcrit$, the scale beyond which larger batches produce
diminishing statistical returns.

A substantial body of work has shown that $\Bcrit$ is not fixed and typically grows during training~\cite{gns,cbs,cbs_revisited}.
Consequently, the statistically preferred $\Bg$ should also evolve over the course of a run. In practice, however, GNS is only a proxy for $\Bcrit$: converting the raw noise-to-signal ratio into a usable batch size requires a calibration factor that is not known a priori and can vary across models and training stages~\cite{gns, cbs_revisited}. For our purposes, the key takeaway is simply that the statistically efficient batch regime changes during training.

\subsection{The Batch-Parallelism Coupling}
\label{subsec:coupling}

The hardware side of the problem is captured by system throughput, which
depends on both the parallel strategy and the batch decomposition:
\begin{equation}
  \mathrm{T}(S, \Bg, \Bm, H)
  = \frac{\Bg}{T_{\mathrm{iter}}(S, \Bg, \Bm, H)},
  \label{eq:throughput_background}
\end{equation}
where $H$ denotes the hardware configuration and $T_{\mathrm{iter}}$ is
the iteration time. For a fixed model and cluster, the throughput-optimal strategy is therefore
\begin{equation}
  S^\star_{(\Bg,\Bm)} = \arg\max_S \mathrm{T}(S, \Bg, \Bm, H).
  \label{eq:s_star_background}
\end{equation}
The key empirical observation of this work is that $S^\star$ need not
be fixed: as shown in \autoref{fig:throughput_coupling}, the
throughput-optimal strategy may shift markedly as $\Bg$ changes. When
$\Bg$ is small, data parallelism is often underutilized because each
replica receives too little work, so TP- or PP-heavier strategies can
achieve higher utilization despite their higher communication cost. As
$\Bg$ grows, DP-dominant strategies become more attractive because
synchronization is amortized across more samples, allowing the system
to sustain larger effective workloads.

Combining this observation with the fact that the statistically
efficient batch size changes over training yields the core motivation
for \sys{}. If $\Bg$ should grow as training progresses, and if the
throughput-optimal parallel strategy depends on the current batch size,
then the best parallel strategy should evolve as well. Any method that
fixes $S$ while adapting $\Bg$, or optimizes $S$ for a fixed $\Bg$, is
therefore solving only one side of a coupled problem and will operate
suboptimally for part of the run.

\section{Goodput-Driven Co-Optimization}
\label{sec:goodput}

The previous section established the coupling problem: the statistically preferred batch size changes during training, and the throughput-optimal 3D parallel strategy changes with it. The remaining question is what objective should govern the joint choice of $(S,\Bg,\Bm)$. We argue that an appropriate objective is \emph{Goodput}, because it directly measures convergence per unit time rather than optimizing statistical efficiency or hardware throughput in isolation. We then refine this objective for Adam-based LLM training and show why it is less sensitive to GNS scaling-factor uncertainty than CBS-only batch selection.

\subsection{From Per-Sample Efficiency to Goodput}
\label{subsec:goodput_metric}

Neither statistical efficiency nor throughput is sufficient on its own. Statistical efficiency captures optimization progress per processed sample but ignores execution time; throughput captures execution speed but ignores whether the chosen batch regime is statistically efficient. What matters in practice is convergence per unit wall-clock time.

Pollux~\cite{adaptdl} introduced \emph{Goodput} to capture this trade-off in multi-job, data-parallel cluster scheduling.
In that setting, the system co-adapts batch size and resource
allocation. In our setting, the job typically runs on a fixed set
of GPUs, so the key decision is different: not how many
resources to allocate, but how to configure the available
resources. Goodput is therefore an even more natural objective
for LLM training, because it lets the system evaluate each
candidate execution tuple $(S,\Bg,\Bm)$ jointly:
\begin{equation}
  \mathrm{Goodput}_t(S,\Bg,\Bm,H)
  = \mathrm{T}(S,\Bg,\Bm,H)\times \SE_t(\Bg).
  \label{eq:goodput}
\end{equation}

To make Eq.~\ref{eq:goodput} usable online, we need a model
for $\SE_t(\Bg)$ from the current GNS measurement.
Following the GNS-based scaling law used by
Pollux~\cite{adaptdl}, the optimization-side effect of batch
size can be expressed as a diminishing-returns curve. Since
our throughput term is measured in samples per second, we
use the corresponding \emph{per-sample} efficiency:
\begin{equation}
  \SE_t(\Bg)
  = \frac{1+\GNS_t}{\Bg+\GNS_t}.
  \label{eq:se_gns}
\end{equation}
This expression has the expected behavior. When
$\Bg \ll \GNS_t$, the batch is in the noise-dominated regime
and the per-sample efficiency is close to its maximum. When
$\Bg \gg \GNS_t$, additional samples provide diminishing
returns and $\SE_t(\Bg)$ decays approximately as $1/\Bg$.
Goodput therefore captures the central trade-off of large-model
training: small batches are statistically attractive but may run
poorly on hardware, while large batches can run quickly but
waste samples.

Most importantly, Goodput does more than output a single
target batch size. It induces a ranking over the full space of
valid configurations:
\[
  (S_{t}^\star,\Bg{}_{t}^\star,\Bm{}_{t}^\star) = \arg\max_{(S,\Bg,\Bm)\in\mathcal{C}(H)}
  \mathrm{Goodput}_t(S,\Bg,\Bm,H),
\]
where $\mathcal{C}(H)$ is the set of configurations that are
valid on hardware $H$ under memory and divisibility
constraints. This is the key difference from CBS-style
selection. CBS chooses a batch size from a statistical signal
and leaves the system's choice to a later stage, whereas
Goodput makes the batch-size and parallelism decisions
jointly.

\subsection{LR-Aware Goodput}
\label{subsec:lr_goodput}

The objective above still inherits an assumption from prior
work: it treats the statistical efficiency term as the only
optimization-side effect of changing batch size. This is appropriate
when learning-rate adaptation is handled outside the Goodput expression,
as in Pollux's plug-in LR scaler, which uses AdaScale for SGD
workloads~\cite{adascale,adaptdl}. Modern LLM pre-training, however,
typically uses Adam-based optimizers with explicit
batch-size-dependent learning-rate scaling, most commonly the square-root rule
$\eta(\Bg)\propto \sqrt{\Bg}$~\cite{sqrt_lr}. Under this
regime, a larger batch can improve convergence not only by
reducing gradient noise, but also by enabling a larger learning
rate.

To account for this, we model convergence rate as
\emph{step rate} times \emph{per-step progress}. The step rate
is $\mathrm{T}(S,\Bg,\Bm,H)/\Bg$, because
$\mathrm{T}(S,\Bg,\Bm,H)$ is measured in samples per second.
The per-step progress scales with the number of samples in the
step, the per-sample efficiency, and the learning rate. Up to a
reference-dependent constant, this gives
\begin{equation}
\begin{aligned}
\mathrm{Goodput}^{\mathrm{LR}}_t(S,\Bg,\Bm,H)
&= \mathrm{T}(S,\Bg,\Bm,H)\,\SE_t(\Bg) \\
&\quad \times \frac{\eta(\Bg)}{\eta(\Bg^{\mathrm{ref}})} .
\end{aligned}
\label{eq:goodput_lr_general}
\end{equation}
For Adam with square-root batch scaling,
\[
  \eta(\Bg)
  = \eta(\Bg^{\mathrm{ref}})
    \sqrt{\frac{\Bg}{\Bg^{\mathrm{ref}}}},
\]
so the candidate ranking is equivalently obtained by
maximizing
\begin{equation}
\begin{aligned}
\mathrm{Goodput}^{\mathrm{LR}}_t(S,\Bg,\Bm,H)
&\propto \mathrm{T}(S,\Bg,\Bm,H)\,\SE_t(\Bg) \\
&\quad \times \sqrt{\Bg} .
\end{aligned}
\label{eq:goodput_lr}
\end{equation}
The constant factor depending on $\Bg^{\mathrm{ref}}$ can be
dropped because it is shared by all candidates. Without this
correction, the objective systematically undervalues larger
batches in training recipes that increase the
learning rate with batch size.

This correction also exposes a structural limitation of
CBS-only selection. GNS captures a property of the gradient
distribution; it does not encode how the optimizer rescales the
learning rate as $\Bg$ changes. A CBS-style rule that maps GNS
directly to a target batch size must therefore absorb two
distinct effects into the same calibration constant: the
GNS-to-CBS mismatch and the LR-scaling effect. Goodput
keeps them separate. The GNS estimate shapes the
diminishing-returns term $\SE_t(\Bg)$, while the optimizer
rule appears explicitly as the LR factor in
Eq.~\ref{eq:goodput_lr}. This makes the objective both more
interpretable and more faithful to how LLM training recipes
are actually executed.

\subsection{Goodput vs.\ CBS Under Scaling Factor Uncertainty}
\label{subsec:goodput_vs_cbs}

A practical concern is that GNS requires a calibration factor to translate the raw noise-to-signal ratio into a usable batch size, and this factor is not known a priori~\cite{gns,cbs_revisited,layernorm}. CBS-style methods treat the scaled GNS directly as the batch-size decision, so any calibration error passes through linearly. Goodput is less sensitive to this error for three reasons: it models statistical efficiency as a continuous curve rather than a single CBS number, the throughput surface provides an independent constraint unaffected by GNS error, and the decision is made over the full tuple $(S,\Bg,\Bm)$ so the selected batch size already accounts for hardware execution. Appendix~\ref{appendix:goodput_vs_cbs} makes this intuition explicit in a simplified model: if the GNS-derived critical-batch estimate is off by a multiplicative factor $c$, the Goodput-selected batch size scales with $\sqrt{c}$ rather than $c$. Although real throughput is discrete and the exact error reduction does not hold universally, the same intuition applies: Goodput is anchored by both the statistical signal and the hardware surface, making it less sensitive to GNS miscalibration.

Empirical CBS methods avoid this calibration problem by measuring the
batch-size threshold directly. For example, branching-based methods
launch several short training branches from a checkpoint, each with a
different batch size and learning-rate scaling rule, and select the
largest batch size whose loss remains close to smaller-batch branches
after a fixed token window~\cite{cbs,cbs_revisited}. This provides a
more direct statistical target than raw GNS, but it requires additional
training runs at each measurement point and returns only a $\Bg$ target.
It does not decide which 3D parallel layout or micro-batch decomposition
maximizes wall-clock progress, so it is complementary to our Goodput
controller rather than a replacement.

\section{\sys{} System Design}
\begin{figure}
    \centering
    \includegraphics[width=\linewidth]{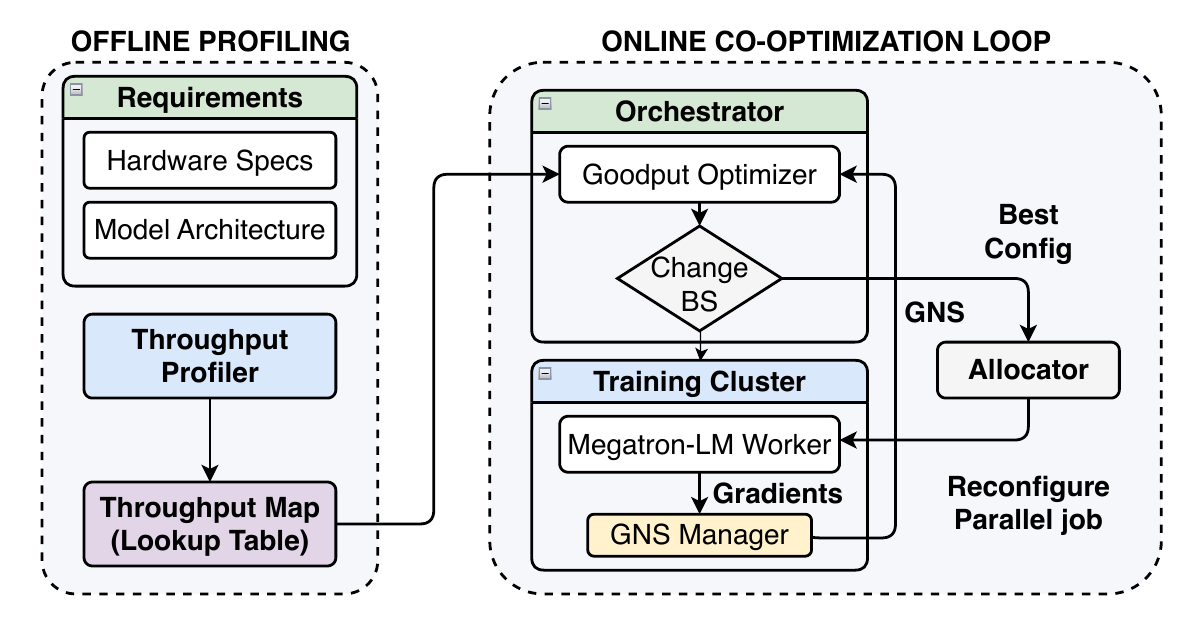}
    \caption{\sys{} system architecture. An offline throughput profile and online GNS measurements feed the orchestrator, which evaluates Goodput across candidate $(S, \Bg, \Bm)$ configurations and triggers batch size or parallelism changes.}
    \label{fig:system_architecture}
\end{figure}

\label{sec:design}

\sys{} minimizes wall-clock time to convergence by co-adapting the parallelism strategy, batch size, and micro-batch size throughout a single training run. It continuously selects the $(S, \Bg, \Bm)$ configuration that maximizes Goodput.

\subsection{Overview}
\label{subsec:overview}

\autoref{fig:system_architecture} shows the \sys{} control loop. The system has two parts: the training process and an out-of-process orchestrator. The training process runs the forward-backward-optimizer loop and contains a GNS manager that estimates gradient noise scale under 3D parallelism (\S\ref{subsec:gns_estimation}). Our orchestrator receives these estimates periodically and combines them with throughput measurements to evaluate Goodput across all candidate $(S, \Bg, \Bm)$ configurations (\S\ref{subsec:orchestrator}). When a better configuration exists, it triggers one of two reconfiguration paths, a batch size change or a parallelism change (\S\ref{subsec:reconfiguration}).

This separation keeps the GPU critical path simple. The
training loop only computes the statistics needed for GNS,
while the search over candidate configurations and the switch
policy live in the orchestrator. The offline throughput table
lets the orchestrator rank candidate configurations at runtime
by combining table lookups with the current GNS estimate.

\subsection{Online GNS Estimation Under 3D Parallelism}
\label{subsec:gns_estimation}

\begin{figure}
    \centering
    \includegraphics[width=\linewidth]{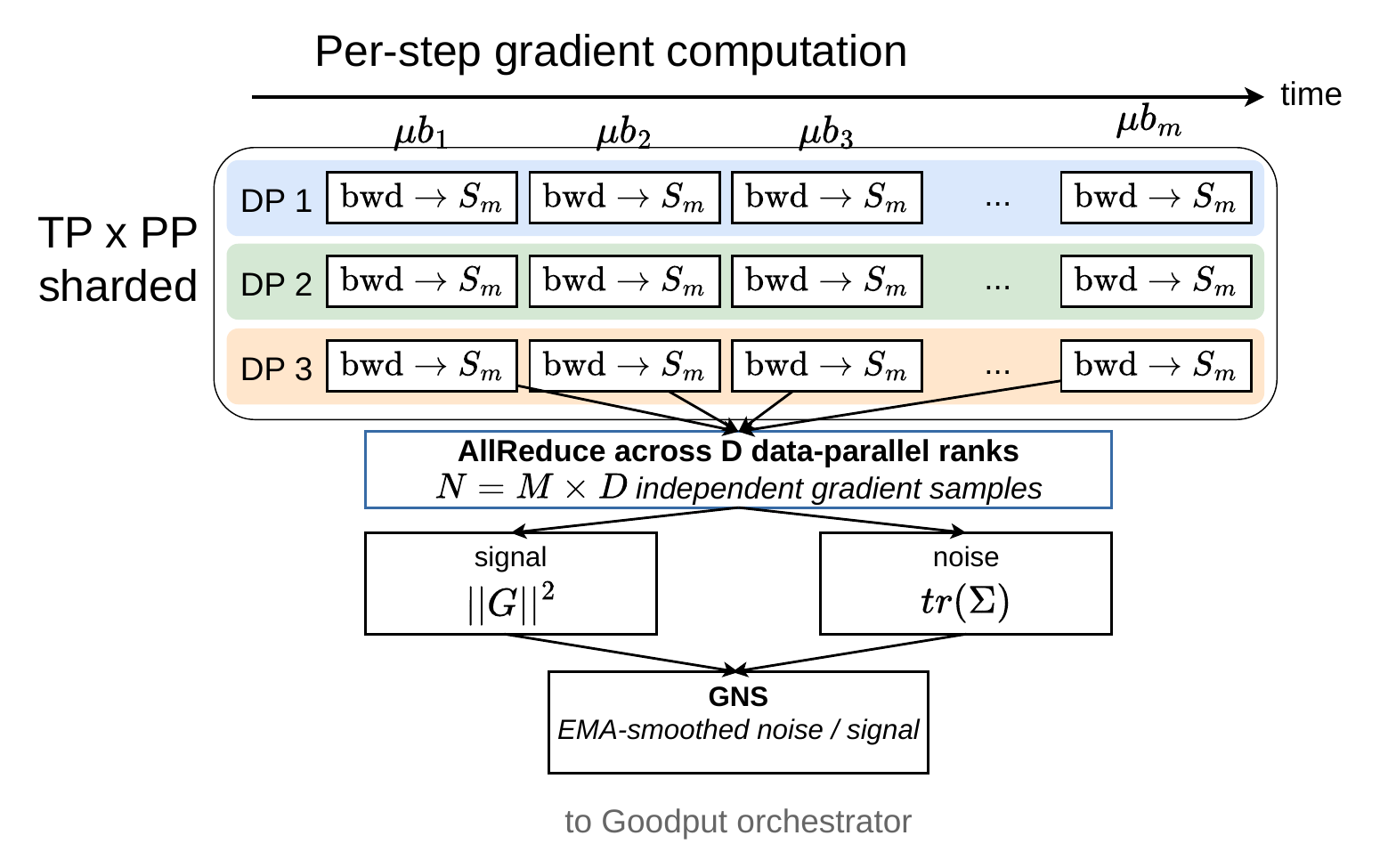}
    \caption{GNS estimation under 3D parallelism. Per-microbatch gradient norms and the all-reduced mean gradient yield the signal $\lVert G \rVert^2$ and noise $\mathrm{tr}(\Sigma)$ used for Goodput evaluation.}
    \label{fig:gns_diagram}
\end{figure}

\autoref{fig:gns_diagram} summarizes how \sys{} estimates GNS
under 3D parallelism. To select the best batch size, \sys{} needs
the gradient noise scale $\GNS$ (Equation~\ref{eq:gns_background}). Prior
estimators such as Pollux~\cite{adaptdl} assume pure data
parallelism: every worker processes a different mini-batch, so
per-worker gradient norms directly provide independent noise
samples. This assumption breaks under 3D parallelism. With
tensor parallelism, no single rank holds a full gradient. With
pipeline parallelism, each rank only sees a subset of layers.
A usable estimator must therefore recover the statistical signal
without ever materializing a full, pre-reduction gradient on a
single worker.

\sys{} addresses this by treating gradient-accumulation micro-batches as the independent stochastic samples. This matches how large-model training is already executed. During backpropagation, we capture the squared gradient norm for each micro-batch, then combine these local statistics with the synchronized mean gradient to recover both $\mathrm{tr}(\Sigma)$ and $\lVert G \rVert^2$ for the current step. This design works under arbitrary $(d,t,p)$ configurations because it never requires any rank to hold the full unsharded gradient.

The estimator also addresses two practical issues. First, we
normalize by actual token counts so that the statistics remain
consistent under variable-length sequences. Second, because
both the signal and noise estimates are inherently volatile, we
smooth them with an exponential moving average before they
are used by the orchestrator. The resulting estimator is
3D-parallel-aware, incurs near-zero overhead, and produces a
single GNS stream that is valid under whichever parallelism
strategy is currently active. Algorithm~\ref{alg:gns}
shows the procedure.

\begin{algorithm}[t]
\begin{algorithmic}[1]
\State \textbf{Input:} Model parameters $\theta$, micro-batch count $M$, DP size $D$
\State \textbf{Output:} Updated GNS estimates $\hat{\phi}_{\mathrm{sqr}}$, $\hat{\phi}_{\mathrm{var}}$
\Statex
\For{each micro-batch $m = 1, \ldots, M$}
    \State Run backward pass
    \State $s_m \gets \sum_p \lVert \nabla_p \mathcal{L}_m \rVert^2$ \Comment{Local squared norm}
\EndFor
\Statex
\State $N \gets M \times D$ \Comment{Total independent samples}
\State $\bar{s} \gets \frac{1}{N} \operatorname{AllReduce}\bigl(\sum_m s_m\bigr)$ \Comment{Average of squared norms}
\State $g_{\mathrm{total}} \gets \operatorname{AllReduce}\bigl(\nabla \mathcal{L}\bigr)$ \Comment{Standard gradient sync}
\State $\bar{g}^2 \gets \lVert g_{\mathrm{total}} \rVert^2$ \Comment{Squared norm of mean gradient}
\Statex
\State $\lVert G \rVert^2 \gets \frac{N \cdot \bar{g}^2 - \bar{s}}{N - 1}$ \Comment{True gradient signal}
\State $\mathrm{tr}(\Sigma) \gets \frac{(\bar{s} - \bar{g}^2) \cdot \Bg}{N - 1}$ \Comment{Gradient noise}
\Statex
\State $\hat{\phi}_{\mathrm{sqr}} \gets \alpha \cdot \hat{\phi}_{\mathrm{sqr}} + (1 - \alpha) \cdot \lVert G \rVert^2$ \Comment{EMA smoothing}
\State $\hat{\phi}_{\mathrm{var}} \gets \alpha \cdot \hat{\phi}_{\mathrm{var}} + (1 - \alpha) \cdot \mathrm{tr}(\Sigma)$
\Statex
\State $\phi \gets \hat{\phi}_{\mathrm{var}} \,/\, \hat{\phi}_{\mathrm{sqr}}$ \Comment{Gradient noise scale (GNS)}
\end{algorithmic}
\caption{Online GNS estimation under 3D parallelism.}
\label{alg:gns}
\end{algorithm}

The result is a GNS estimate under any $(d, t, p)$ configuration. The total number of independent samples is $N = M \times D$ (micro-batches times DP ranks), so even with small DP size, gradient accumulation provides enough samples.

\subsection{The Goodput Orchestrator}
\label{subsec:orchestrator}

Our orchestrator periodically receives GNS estimates from the training process. For each candidate $(S', \Bg', \Bm')$ in the throughput table, it computes statistical efficiency from the current GNS estimate, looks up the throughput, and applies the LR-aware Goodput formula from \S\ref{subsec:lr_goodput}. It then selects the candidate with the highest Goodput.

Algorithm~\ref{alg:orchestrator} shows the decision loop. The orchestrator can issue three commands. If the current configuration is already best, it does nothing (No-Op). If the best candidate has the same parallelism but a different batch size, it issues a batch size update (Scale-BS). If the best candidate requires a different parallelism strategy, it triggers an online reconfiguration (Reconfigure).

The orchestrator uses two mechanisms to avoid unnecessary or harmful switches. First, GNS estimates fluctuate because they are based on finite gradient samples. Two candidates may have similar Goodput, and noise alone can make either one appear better from one iteration to the next. A switching margin $\epsilon$ suppresses changes unless the best candidate exceeds the current Goodput by at least $\epsilon$. If the margin is too small, the orchestrator oscillates between similar configurations. If it is too large, the orchestrator reacts slowly to real changes in the training dynamics. We find that $\epsilon = 10\%$ works well across our experiments (\S\ref{sec:evaluation}).

Second, parallelism changes pause training while \sys{}
reconstructs the runtime and reshards the persistent training
state. The orchestrator needs to decide whether the
projected Goodput gain from a new parallelism strategy is
worth the one-time reconfiguration cost. We model this with a
reallocation factor
$T_{\mathrm{useful}} / (T_{\mathrm{elapsed}} + c_{\mathrm{reconfig}})$,
where $T_{\mathrm{elapsed}}$ is the elapsed wall-clock time,
$T_{\mathrm{useful}}$ subtracts any prior reconfiguration overhead,
and $c_{\mathrm{reconfig}}$ is the measured cost of the online
reconfiguration. The factor represents the fraction of total
wall-clock time spent on useful training if we switch now. Early in
training, $T_{\mathrm{useful}}$ is small and the factor is low, so the
orchestrator avoids expensive
parallelism changes unless the gain is large. As useful training time
accumulates, the factor approaches 1, and the same
one-time pause becomes easier to amortize.

\begin{algorithm}[t]
\begin{algorithmic}[1]
\State \textbf{Input:} GNS estimates $(\hat{\phi}_{\mathrm{sqr}}, \hat{\phi}_{\mathrm{var}})$, throughput table $\mathcal{T}$, current config $(S, \Bg, \Bm)$, margin $\epsilon$, times $T_{\mathrm{elapsed}}, T_{\mathrm{useful}}$
\State \textbf{Output:} Command $\in$ \{No-Op, Scale-BS, Reconfigure\}
\Statex
\For{each candidate $(S', \Bg', \Bm')$ in $\mathcal{T}$}
    \State $\SE \gets \textsc{StatEff}(\Bg', \hat{\phi}_{\mathrm{sqr}}, \hat{\phi}_{\mathrm{var}})$
    \State $G \gets \mathcal{T}(S', \Bg', \Bm') \times \SE \times \sqrt{\Bg'}$ \Comment{LR-aware Goodput}
    \If{$S' \neq S$}
        \State $G \gets G \times T_{\mathrm{useful}} \,/\, (T_{\mathrm{elapsed}} + c_{\mathrm{reconfig}})$ \Comment{Penalize reconfiguration cost}
    \EndIf
\EndFor
\Statex
\State $(S^*, \Bg^*, \Bm^*) \gets \arg\max G$
\State $G_{\mathrm{cur}} \gets$ Goodput of current config
\If{$(G^* - G_{\mathrm{cur}}) / G_{\mathrm{cur}} < \epsilon$}
    \State \Return No-Op
\ElsIf{$S^* = S$}
    \State \Return Scale-BS$(\Bg^*, \Bm^*)$
\Else
    \State \Return Reconfigure$(S^*, \Bg^*, \Bm^*)$
\EndIf
\end{algorithmic}
\caption{Orchestrator decision loop.}
\label{alg:orchestrator}
\end{algorithm}

\subsection{Throughput Profiling}
\label{subsec:throughput_table}

The throughput table is generated by offline benchmarking of all
memory-feasible $(S, \Bg, \Bm)$ configurations for a given
model-hardware pair; configurations that exceed GPU memory are pruned.
This direct-measurement approach is simple and captures hardware
effects, but requires a one-time profiling pass per model-hardware
pair. More automated systems use analytical or simulator-based cost
models to reduce this cost~\cite{alpa,galvatron,flexflow}. Our contribution is
not the profiler itself, but feeding the resulting throughput table into
the Goodput optimizer so that throughput enters the batch-size decision
continuously. Details are provided in
Appendix~\ref{appendix:throughput_profiling}.

\subsection{Reconfiguration Mechanisms}
\label{subsec:reconfiguration}

When our orchestrator selects a new configuration, \sys{} must apply it without losing training state. The cost depends on what changed. We support two reconfiguration paths.

\paragraph{Batch size changes.}
Changing $\Bg$ and $\Bm$ does not require a restart. We broadcast the new values to all ranks, which rebuild their micro-batch calculators and update batching state to match the new batch dimensions. We scale the learning rate following the square-root rule for Adam~\cite{sqrt_lr},
\begin{equation}
  \eta' = \eta \cdot \sqrt{\frac{\Bg'}{\Bg}},
  \label{eq:lr_scaling}
\end{equation}
and training resumes at the next iteration. This is the common case. Since the critical batch size grows during training, our orchestrator adjusts $\Bg$ frequently through this path.

\paragraph{Parallelism changes.}
Changing the parallel strategy $S = (d, t, p)$ is more
involved because it changes both the communication topology
and the shard layout of persistent training state. After such a
change, training continues under the new layout, so \sys{} must
reshard not only the model weights but also the optimizer state before
training can resume.

We perform this as online state resharding at an optimizer-step
boundary, avoiding a full checkpoint-restart. The system pauses training
between steps, extracts the source shards of the current model and
optimizer state into CPU host memory, releases the current GPU-resident
model and optimizer state, reconstructs the process groups for the
target topology, rebuilds the model and optimizer under those groups,
and loads the staged state into the target shard layout. Because source
state is staged in CPU memory before target state is materialized on
GPU, the procedure does not require any GPU to hold both layouts at
once; it stays within the memory footprint of a single valid
configuration, assuming both the source and target configurations fit
individually. Appendix~\ref{appendix:resharding} details the pipeline,
and \S\ref{subsec:overhead} compares its latency against full
checkpoint-restart.

Parallelism changes are rare. In our experiments, each run triggers only one or two, but each one unlocks a new throughput regime that persists for a significant fraction of the remaining training. The cost-benefit analysis in Algorithm~\ref{alg:orchestrator} ensures that these changes occur when the projected long-term Goodput gain justifies the one-time pause
(\S\ref{subsec:overhead}).

\section{Implementation}
\label{sec:implementation}

We implement \sys{} as a fork of
Megatron-LM~\cite{megatron-lm,large_megatron-lm}. We use
Megatron-Core~0.14.0 on NVIDIA and 0.15.0 on AMD hardware, with
PyTorch~\cite{pytorch}, AdamW~\cite{loshchilov2018decoupled}, and BF16 mixed
precision. The main additions are a hook-based GNS estimator that
captures per-microbatch gradient norms during backpropagation
(\S\ref{subsec:gns_estimation}), an adaptive batch path that changes
$\Bg$ and $\Bm$ between optimizer steps without restarting or
resetting dataset traversal, an online resharding pipeline that
reconfigures the 3D-parallel topology in-process by staging state in
host memory (Appendix~\ref{appendix:resharding}), an offline throughput
lookup table indexed by $(S,\Bg,\Bm)$, and an out-of-process
orchestrator connected to rank~0 via WebSocket. Only rank~0
communicates with the orchestrator; commands are broadcast so all
workers transition atomically at the same optimizer-step boundary.
Detailed descriptions of each component are provided in
Appendix~\ref{appendix:implementation}.

\section{Evaluation}
\label{sec:evaluation}

\subsection{Experimental Setup}
\label{subsec:setup}

\begin{figure}[t]
    \centering
    \includegraphics[width=\linewidth]{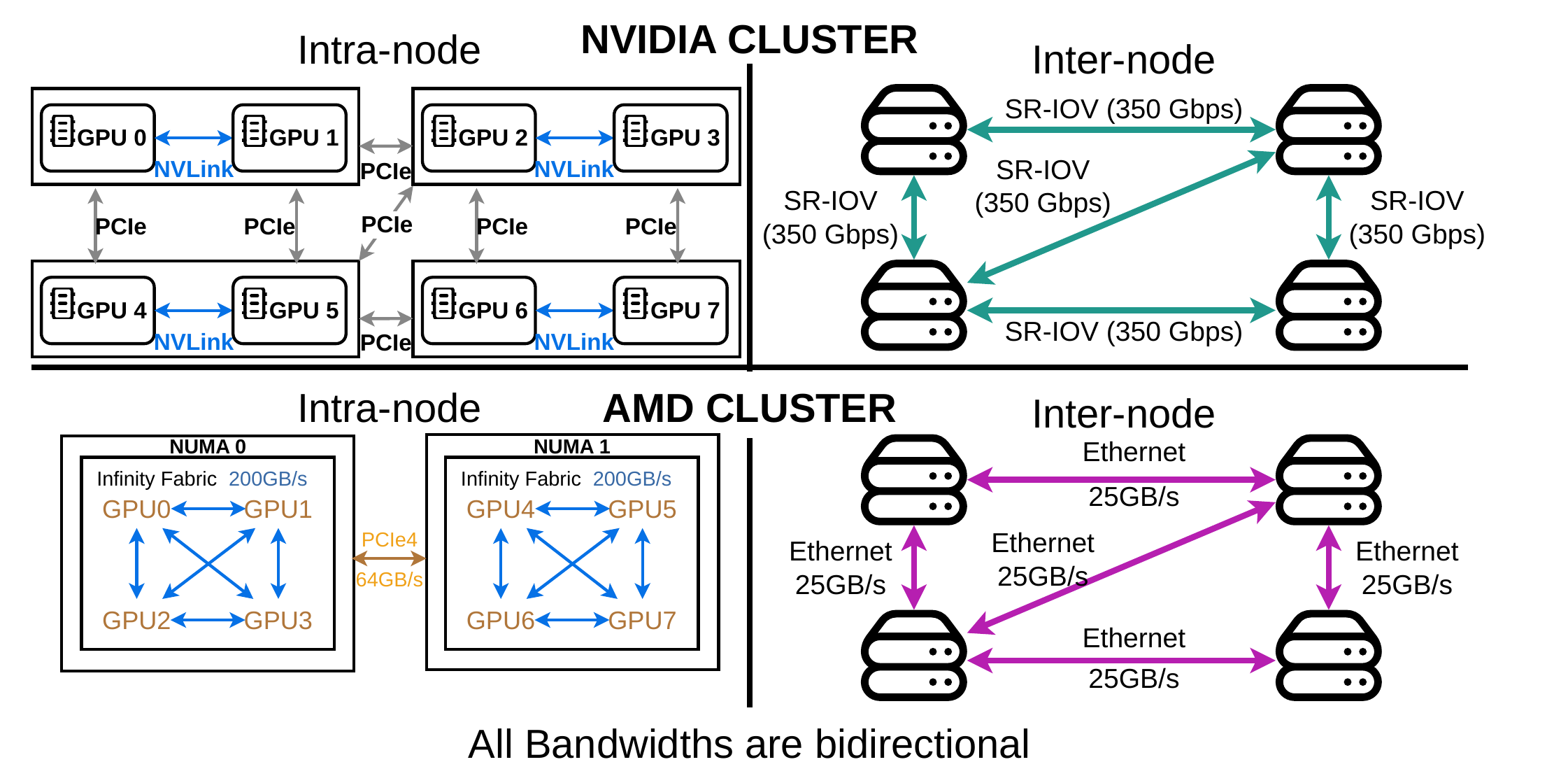}
    \caption{Interconnect topology of our two evaluation clusters.
    In both clusters, GPUs within a node are not uniformly connected.
    NVLink (NVIDIA) and Infinity Fabric (AMD) only cover subsets of
    GPUs, so tensor parallelism across all 8 GPUs must cross slower
    PCIe links.}
    \label{fig:topology}
\end{figure}

\paragraph{Hardware.}
We evaluate \sys{} on two clusters with 8 GPUs per node
(\autoref{fig:topology}). Our NVIDIA cluster uses H100 GPUs connected
in NVLink pairs, with PCIe between pairs and SR-IOV (350~Gbps) between
nodes. Our AMD cluster uses MI210 GPUs with Infinity Fabric
(200~GB/s) within each NUMA domain, PCIe4 (64~GB/s) across NUMA
domains, and Ethernet (25~GB/s) between nodes. In both clusters,
high-bandwidth links cover only subsets of GPUs within a node;
configurations with TP${}>{}$2 on the NVIDIA cluster or TP${}>{}$4
on the AMD cluster must traverse slower interconnects.

\paragraph{Models and data.}
We pre-train four transformer models: LLaMA-3.2-3B~\cite{llama32} on 1 node
(8 H100), LLaMA-2-13B~\cite{llama2} on 2 nodes (16 H100), Qwen-2.5-32B~\cite{qwen2.5} on 4 nodes
(32 H100), and LLaMA-2-7B~\cite{llama2} on 4 nodes (32 MI210). All models are
trained on the WikiText-103 dataset~\cite{wikitext103} with a sequence
length of 2{,}048 tokens, using each model's own pre-trained
tokenizer. Each experiment runs for a fixed token budget:
328M tokens for the 3B and 13B configurations, 123M tokens for 32B,
and 215M tokens for 7B. Because the budget is fixed in tokens, runs
with larger batch sizes complete in fewer wall-clock minutes and
appear shorter in the time-axis plots.

\paragraph{Training hyperparameters.}
All experiments use BF16 and AdamW~\cite{loshchilov2018decoupled} with $\beta_1 = 0.9$
and $\beta_2 = 0.95$, following standard LLM pre-training
practice~\cite{llama,opt}. All models start at $\Bg = 16$.
Base learning rates, chosen by a short sweep at this batch size, are
$2 \times 10^{-4}$ for 3B and 7B, $1 \times 10^{-4}$ for 13B, and
$7 \times 10^{-5}$ for 32B. For all methods, including static-GBS
baselines, the learning rate scales as $\sqrt{\Bg / 16}$
following the square-root Adam rule~\cite{sqrt_lr}. We warm up
linearly over the first 8M tokens (about 4{,}000 samples), then keep
the rate constant for the rest of the budget.

\paragraph{GNS estimation.}
We smooth the gradient noise scale with an exponential moving average,
using $\alpha = 0.95$ for the first 8M tokens and $\alpha = 0.99$
thereafter. The lower initial value reduces bias from the noisiest
early measurements; the higher later value improves stability. All
GNS-based methods (CBS and \sys{}) share a calibration factor
$c = 2.0$ on the variance term $\mathrm{tr}(\Sigma)$ in the GNS ratio
(\autoref{eq:gns_background}, \S\ref{subsec:se_cbs}), which scales the
estimated $\Bcrit$ used by Goodput. Automatically determining $c$ is
beyond the scope of this evaluation, and a linear correction is only an
approximation of the true GNS-to-CBS relationship.

\paragraph{\sys{} configuration.}
\sys{} switches only when the best candidate exceeds the current
configuration's Goodput by at least 10\% (the \emph{switching
margin}), which suppresses oscillation between similarly ranked
candidates. To limit optimizer shock, the global batch size may grow
by at most $2\times$ in a single step.

\paragraph{Baselines.}
We compare against two baseline families. \emph{Static-GBS} baselines
fix the global batch size and use the throughput-optimal parallelism
strategy and micro-batch size for that batch, as determined by
profiling. \emph{CBS} baselines adapt $\Bg$ online by choosing the
candidate batch size closest to the GNS-estimated critical batch size,
using the same GNS estimator and calibration factor as \sys{}, but
keep parallelism and micro-batch size fixed. For each experiment, we
run two CBS variants: a \emph{pessimistic} one optimized for the
initial $\Bg = 16$, and an \emph{optimistic} one tuned for the
high-$\Bg$ regime that dominates most of training. Together, these two
variants cover the fixed-parallelism choices a practitioner might make
without co-adapting parallelism.

\paragraph{Throughput profiling and search space.}
The throughput profile enumerates all memory-feasible
$(S, \Bg, \Bm)$ combinations and retains only the fastest
micro-batch size for each $(S, \Bg)$ pair. These measured throughput
profiles are the same ones shown in \autoref{fig:throughput_coupling}
and serve as the lookup table used by the adaptive policies. Our
decision space covers data, tensor, and pipeline parallelism; other
dimensions used in LLM training, such as ZeRO-style optimizer
sharding~\cite{zero}, context parallelism, and sequence parallelism,
could be added as additional search dimensions and are left as future
work. All configurations use replicated optimizer state across
data-parallel ranks. In adaptive runs (\sys{} and CBS), decisions are
made from the profiled table rather than live measurements, so
slight performance fluctuations do not affect the decision-making controller.

\subsection{End-to-End Convergence}
\label{subsec:e2e}

\begin{figure*}[t]
    \centering
    \includegraphics[width=0.92\textwidth]{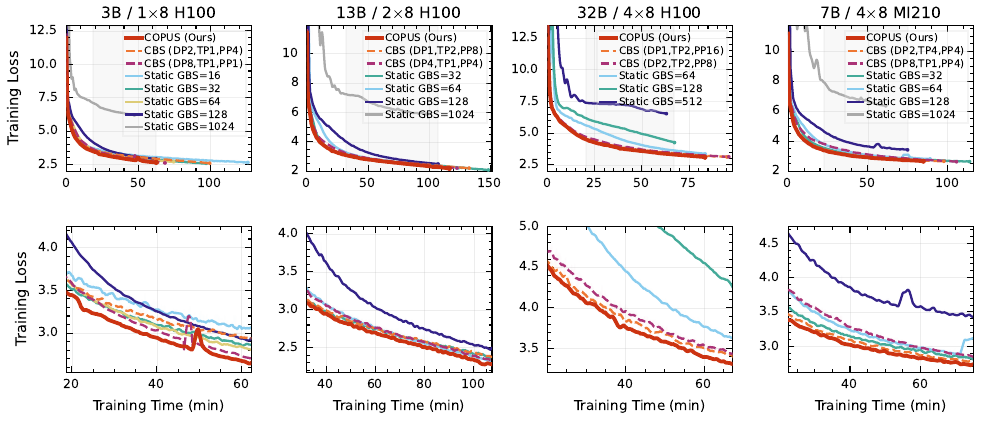}
    \caption{Training loss vs.\ training time across all four configurations. Top row: full training view. Bottom row: zoomed convergence region (shaded area in top row). Thick lines show Savitzky-Golay smoothed curves (2\,min window, 3rd-order); faint lines show raw data. Static baselines use the throughput-optimal parallelism strategy and micro-batch size for their respective batch sizes.}
    \label{fig:loss_vs_time}
\end{figure*}

\begin{table*}[t]
\centering
\scriptsize
\setlength{\tabcolsep}{2pt}
\caption{Time to target loss (minutes) and average loss. Bold = best per column; ``$-$'' = target not reached. Avg Loss is the mean loss over the bracketed time window. \textit{Speedup} includes resharding overhead (\autoref{tab:resharding_overhead}); \textit{Ideal} assumes zero overhead. For the \textit{Speedup} and \textit{Ideal} rows, the rightmost column averages across all targets.}
\label{tab:unified_results}
\begin{minipage}[t]{0.48\textwidth}
\centering
\begin{tabular}{@{}l r r r r r r@{}}
\toprule
Method & \multicolumn{5}{c}{Time to target loss (min)} & Avg Loss \\
\cmidrule(lr){2-6}
\textit{\textbf{3B / 1$\times$8 H100}} & \textit{4.0} & \textit{3.5} & \textit{3.2} & \textit{3.0} & \textit{2.75} & \textit{[5,62] min} \\
\cmidrule(lr){1-7}
COPUS (Ours) & \textbf{10.6} & \textbf{18.7} & \textbf{28.7} & \textbf{37.5} & \textbf{54.9} & \textbf{3.2034} \\
CBS (DP2,TP1,PP4) & 11.5 & 22.8 & 37.9 & 55.1 & 83.7 & 3.3978 \\
CBS (DP8,TP1,PP1) & 13.6 & 21.6 & 31.4 & 41.4 & 59.0 & 3.3503 \\
Static GBS=16 & 12.4 & 26.9 & 46.2 & 69.4 & 102.7 & 3.4932 \\
Static GBS=32 & 11.1 & 20.8 & 32.6 & 47.0 & 73.1 & 3.3418 \\
Static GBS=64 & 13.1 & 21.7 & 33.3 & 45.0 & 67.3 & 3.3823 \\
Static GBS=128 & 20.8 & 30.7 & 42.5 & 55.7 & $-$ & 3.7083 \\
Static GBS=1024 & $-$ & $-$ & $-$ & $-$ & $-$ & 6.8589 \\
\cmidrule(lr){1-7}
\textit{Speedup} & +4.3\% & +10.1\% & +8.6\% & +9.4\% & +7.0\% & \textit{+7.9\%} \\
\textit{Ideal} & +8.8\% & +12.5\% & +13.2\% & +12.9\% & +9.5\% & \textit{+11.4\%} \\
\bottomrule
\end{tabular}
\end{minipage}
\hfill
\begin{minipage}[t]{0.48\textwidth}
\centering
\begin{tabular}{@{}l r r r r r r@{}}
\toprule
Method & \multicolumn{5}{c}{Time to target loss (min)} & Avg Loss \\
\cmidrule(lr){2-6}
\textit{\textbf{13B / 2$\times$8 H100}} & \textit{4.0} & \textit{3.5} & \textit{3.0} & \textit{2.75} & \textit{2.5} & \textit{[5,107] min} \\
\cmidrule(lr){1-7}
COPUS (Ours) & \textbf{10.4} & \textbf{18.4} & \textbf{38.0} & \textbf{55.8} & \textbf{81.4} & \textbf{2.8856} \\
CBS (DP1,TP2,PP8) & 10.7 & 19.4 & 39.5 & 59.9 & 88.9 & 2.9450 \\
CBS (DP4,TP1,PP4) & 14.2 & 23.8 & 43.5 & 62.3 & 89.0 & 3.0044 \\
Static GBS=32 & 11.7 & 19.3 & 38.3 & 58.8 & 91.0 & 2.9546 \\
Static GBS=64 & 18.9 & 26.6 & 43.2 & 60.5 & 86.7 & 3.1159 \\
Static GBS=128 & 32.2 & 42.8 & 61.9 & 79.6 & 105.2 & 3.5186 \\
Static GBS=1024 & $-$ & $-$ & $-$ & $-$ & $-$ & 7.3921 \\
\cmidrule(lr){1-7}
\textit{Speedup} & +3.4\% & +4.4\% & +0.7\% & +5.1\% & +6.1\% & \textit{+3.9\%} \\
\textit{Ideal} & +3.4\% & +4.4\% & +2.5\% & +6.3\% & +7.0\% & \textit{+4.7\%} \\
\bottomrule
\end{tabular}
\end{minipage}
\\[4pt]
\begin{minipage}[t]{0.48\textwidth}
\centering
\begin{tabular}{@{}l r r r r r r@{}}
\toprule
\textit{\textbf{32B / 4$\times$8 H100}} & \textit{4.5} & \textit{4.0} & \textit{3.8} & \textit{3.5} & \textit{3.2} & \textit{[5,67] min} \\
\cmidrule(lr){1-7}
COPUS (Ours) & \textbf{20.5} & \textbf{33.0} & \textbf{38.7} & \textbf{55.8} & \textbf{76.5} & \textbf{4.1227} \\
CBS (DP1,TP2,PP16) & 21.3 & 35.4 & 43.5 & 60.5 & 85.7 & 4.1828 \\
CBS (DP2,TP2,PP8) & 24.8 & 38.4 & 46.8 & 62.9 & 87.4 & 4.2831 \\
Static GBS=64 & 39.2 & 51.5 & 59.1 & 74.2 & $-$ & 4.8425 \\
Static GBS=128 & 61.2 & $-$ & $-$ & $-$ & $-$ & 5.5783 \\
Static GBS=512 & $-$ & $-$ & $-$ & $-$ & $-$ & 7.5203 \\
\cmidrule(lr){1-7}
\textit{Speedup} & +3.7\% & +6.8\% & +11.1\% & +7.8\% & +10.7\% & \textit{+8.0\%} \\
\textit{Ideal} & +3.7\% & +6.8\% & +13.1\% & +9.3\% & +11.7\% & \textit{+8.9\%} \\
\bottomrule
\end{tabular}
\end{minipage}
\hfill
\begin{minipage}[t]{0.48\textwidth}
\centering
\begin{tabular}{@{}l r r r r r r@{}}
\toprule
\textit{\textbf{7B / 4$\times$8 MI210}} & \textit{4.0} & \textit{3.5} & \textit{3.2} & \textit{3.0} & \textit{2.8} & \textit{[5,61] min} \\
\cmidrule(lr){1-7}
COPUS (Ours) & 11.1 & \textbf{20.7} & \textbf{31.4} & \textbf{42.1} & \textbf{63.7} & \textbf{3.2983} \\
CBS (DP2,TP4,PP4) & \textbf{11.1} & 21.1 & 32.5 & 46.2 & 71.6 & 3.3471 \\
CBS (DP8,TP1,PP4) & 17.6 & 31.9 & 44.9 & 57.1 & 80.8 & 3.6194 \\
Static GBS=32 & 13.8 & 24.1 & 36.7 & 51.9 & 77.2 & 3.4695 \\
Static GBS=64 & 19.2 & 30.1 & 42.4 & 57.9 & $-$ & 3.6712 \\
Static GBS=128 & 34.9 & 65.6 & $-$ & $-$ & $-$ & 4.3555 \\
Static GBS=1024 & $-$ & $-$ & $-$ & $-$ & $-$ & 7.9621 \\
\cmidrule(lr){1-7}
\textit{Speedup} & -0.5\% & +1.9\% & +3.5\% & +8.9\% & +11.0\% & \textit{+5.0\%} \\
\textit{Ideal} & -0.5\% & +5.8\% & +6.0\% & +10.7\% & +12.1\% & \textit{+6.8\%} \\
\bottomrule
\end{tabular}
\end{minipage}
\end{table*}

\begin{figure}[t]
    \centering
    \includegraphics[width=\columnwidth]{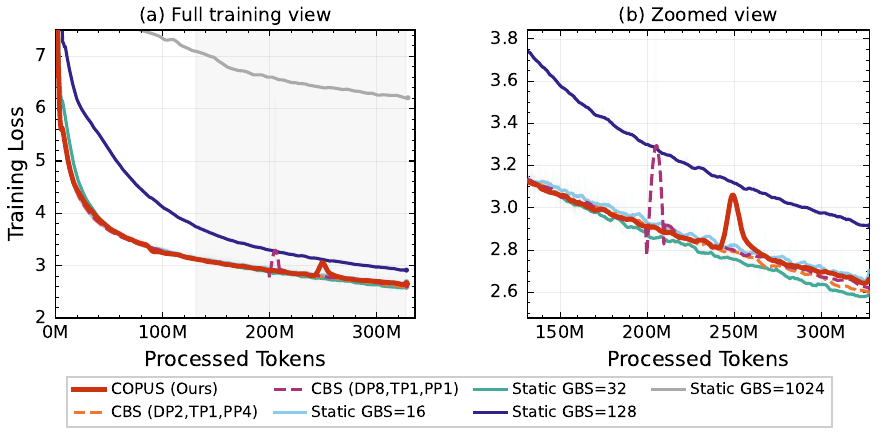}
    \caption{Training loss vs.\ processed tokens (samples $\times$ 2048 sequence length) for the 3B configuration. This view isolates statistical (per-sample) efficiency from throughput: methods with lower loss at the same token count are more statistically efficient.}
    \label{fig:loss_vs_tokens}
\end{figure}

\begin{figure}[t]
    \centering
    \includegraphics[width=\columnwidth]{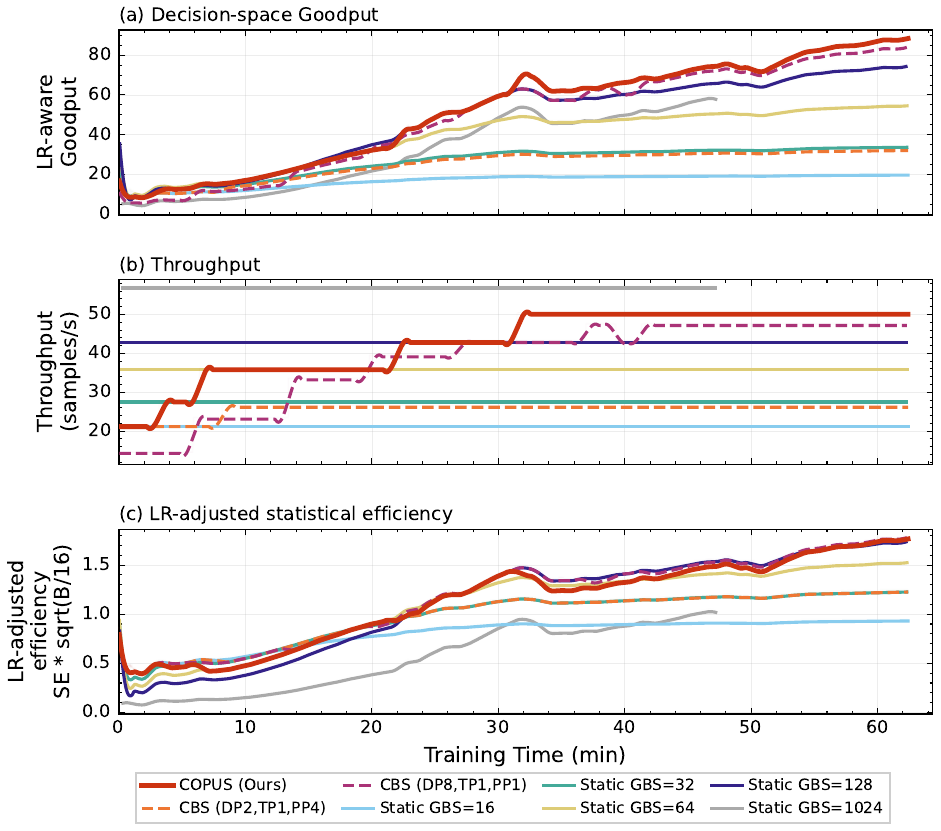}
    \caption{Decision-space Goodput decomposition over training time for the 3B configuration. For each policy, we combine its throughput and batch-size schedule with the GNS trajectory observed by \sys{}, then evaluate the LR-aware objective from \autoref{eq:goodput_lr}. The x-axis is limited to the interval where this \sys{} GNS trajectory is available. (a)~LR-aware Goodput. (b)~Throughput $T(S,\Bg,\Bm)$. (c)~LR-adjusted efficiency $\SE_t(\Bg)\sqrt{\Bg/16}$. The last term can exceed one because it includes square-root learning-rate scaling; the division by $16$ normalizes the display and does not change relative comparisons.}
    \label{fig:goodput_vs_time}
\end{figure}

\begin{figure*}[t]
    \centering
    \small
    \includegraphics[width=0.92\textwidth]{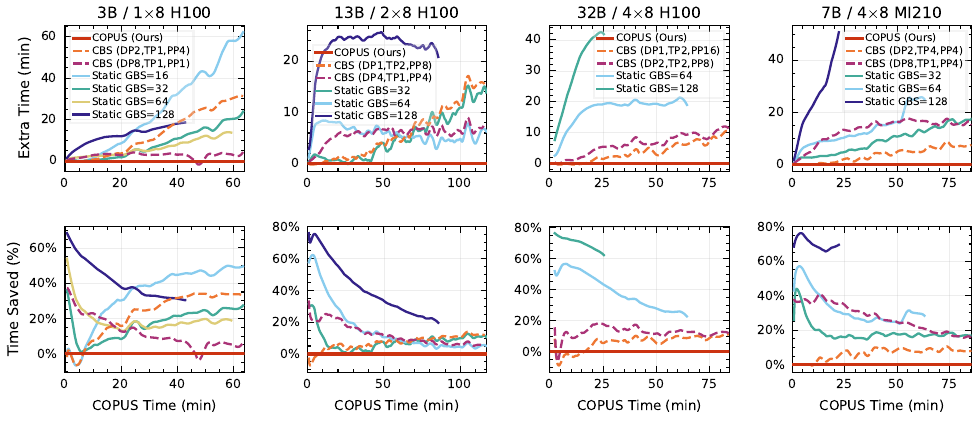}
    \caption{Relative performance of baselines vs.\ \sys{} across all configurations. Top row: extra wall-clock time each baseline needs to reach the same loss as \sys{}. Bottom row: same data as a percentage of the baseline's total time. \sys{} is the zero-line reference. Diverged static baselines are omitted.}
    \label{fig:relative_performance}
\end{figure*}

\sys{} reaches every target loss faster than or comparably to the best
baseline in all four configurations
(\autoref{fig:loss_vs_time}, \autoref{tab:unified_results}).
All speedups include the full cost of online resharding
(\autoref{tab:resharding_overhead}); the ``Ideal'' row in
\autoref{tab:unified_results} shows the additional headroom if this
overhead disappeared entirely. On 3B, \sys{} averages +7.9\% over the
fastest baseline across five loss thresholds, with gains from +4.3\%
to +10.1\%. The 32B configuration averages +8.0\% and peaks at
+11.1\%, where the 4-node topology offers the widest throughput
variation across strategies. The 7B AMD run averages +5.0\% and
reaches +11.0\% at the lowest loss target. The 13B configuration is
more modest at +3.9\% on average because its CBS baselines start from
a strategy that stays near-optimal over most of the traversed batch
range.

Each fastest baseline also assumes an expensive grid search over
parallelism strategies and micro-batch sizes for every static batch
size, and this search cost is not included in the reported baseline
times. Since real training recipes often rely on a single fixed
configuration, these baselines should be viewed as strong, highly tuned
comparisons rather than typical fixed-configuration deployments. This
also means that \sys{}'s margin can narrow later in training when the
strongest fixed baseline was tuned for the high-batch regime that
dominates the later targets, using the same throughput profile.

The speedup depends on whether the adaptive batch trajectory crosses a
throughput reconfiguration boundary
(\autoref{fig:throughput_coupling}). When it does, \sys{} switches to a
higher-throughput strategy; when it does not, \sys{} matches the best
CBS baseline because the Goodput objective reduces to statistical
efficiency optimization when throughput is constant. In this sense,
\sys{} either improves on CBS or matches it.

\autoref{fig:loss_vs_tokens} isolates sample efficiency from
throughput for the 3B experiment. When loss is plotted against
processed tokens rather than wall-clock time, the smallest static
batch (GBS=32) is the most sample-efficient, consistent with critical
batch size theory~\cite{gns}. \sys{} matches the sample efficiency of
the CBS baselines because both use the same GNS-driven batch-size
schedule; it wins on wall-clock time (\autoref{fig:loss_vs_time})
because it co-optimizes throughput. The same loss-versus-token view for
the remaining configurations is provided in
Appendix~\ref{appendix:additional_trajectories}
(\autoref{fig:additional_loss_vs_tokens}).

\subsection{Behavior and Goodput Analysis}
\label{subsec:behavior}

We use the 3B configuration as a detailed case study because it
exhibits the clearest reconfiguration trajectory: three distinct
parallelism strategies over 60 minutes of training. The corresponding
trajectories for the 13B, 32B, and 7B configurations are provided in
Appendix~\ref{appendix:additional_trajectories}
(\autoref{fig:additional_behavior_trajectories}).

As shown in \autoref{fig:behavior_trajectory}, \sys{} jointly evolves
loss, batch size, and parallelism strategy. It starts with
DP2,TP1,PP4, a pipeline-heavy layout suited to the initial
$\Bg = 16$, then moves to DP4,TP1,PP2 at about 3\,min and to fully
data-parallel DP8,TP1,PP1 at about 22\,min as GNS rises and Goodput
favors larger batches. Each transition occurs only when the candidate
exceeds the current Goodput by at least 10\%. The CBS baselines adapt
batch size on a similar schedule, but remain locked to their initial
parallelism. As a result, the DP2,TP1,PP4 baseline loses throughput as
$\Bg$ grows, while the DP8,TP1,PP1 baseline underperforms early when
the batch is too small to saturate pure data parallelism.
Occasional loss spikes appear at batch size transitions and, less
frequently, during steady-state training. We discuss their causes and
connect them to known training dynamics in
Appendix~\ref{appendix:loss_spikes}.

\autoref{fig:goodput_vs_time} decomposes the LR-aware Goodput
objective used by the controller. To compare policies in a common
decision space, the figure evaluates every policy at time $t$ using
the GNS trajectory observed by \sys{}: each policy contributes its
current throughput and batch size, while the statistical-efficiency
term is computed from the same critical batch size. We then apply the
square-root learning-rate factor from \autoref{eq:goodput_lr},
normalized as $\sqrt{\Bg/16}$ for display. This construction asks
which policy the controller objective would prefer in the optimization
states \sys{} actually encountered, rather than mixing different GNS
trajectories across runs.
The corresponding decompositions for the 13B, 32B, and 7B
configurations are provided in
Appendix~\ref{appendix:additional_trajectories}
(\autoref{fig:additional_goodput_decompositions}).

Static GBS=1024 achieves the highest throughput but has a weak
efficiency factor early in training, when such a large batch is
statistically premature. Conversely, the smallest batches retain high
per-sample efficiency but suffer low throughput and receive little
benefit from learning-rate scaling. \sys{} is the only method that
keeps both components high: throughput rises as it shifts to
DP-heavier layouts, while the LR-adjusted efficiency remains
comparable to the strongest CBS baseline because batch size still
tracks the growing critical batch size. Their product shows that
\sys{} maintains the highest Goodput at almost every point in
training, matching the observed convergence gains in
\autoref{fig:loss_vs_time}.

\subsection{Scaling and Generalization}
\label{subsec:generalization}

\autoref{fig:relative_performance} shows the time saved by \sys{}
relative to each baseline across all four configurations. In the 3B
and 7B experiments, \sys{} is consistently faster than the best static
baseline at the evaluated loss targets, saving up to 18.4\% and
19.6\% of training time, respectively. Relative to the best CBS
baseline, the savings reach 13.4\% on 3B and 11.0\% on 7B. The 32B experiment shows the steepest
improvement trajectory: because the 4-node topology has more
parallelism strategies and wider throughput variation across them,
reconfiguration yields a larger benefit. The 13B experiment shows
smaller but consistent gains, reflecting the narrower throughput
spread in its 2-node configuration.

The 7B/MI210 experiment shows that the co-adaptive principle is not
specific to one hardware platform. The MI210 cluster has a different
interconnect hierarchy (\autoref{fig:topology}), yet \sys{}
still benefits from adapting both batch size and parallelism as
training progresses.

\subsection{Reconfiguration Overhead}
\label{subsec:overhead}

\autoref{tab:resharding_overhead} lists every parallelism
reconfiguration in our experiments and compares online resharding
(\S\ref{sec:implementation}) against full checkpoint-restart, which
saves a distributed checkpoint to disk and reloads it under the new
layout. Online resharding reduces latency by 2--16$\times$, while peak
GPU memory never exceeds the footprint of either the source or target
configuration. Because \sys{} changes parallelism only when the
projected Goodput gain persists long enough to amortize the pause, it
does not require sub-second resharding to be effective. Although
30--56\,s is slower than sub-second switching in graph-compiler systems
such as HotSPa~\cite{hotspa}, \sys{} targets a different point:
infrequent, persistent changes within a Megatron-LM pre-training run
that include full optimizer-state resharding. This makes
checkpoint-restart the relevant baseline for our setting.

\begin{table}[H]
\centering
\footnotesize
\setlength{\tabcolsep}{3pt}
\caption{All parallelism reconfigurations that occurred during training. Time indicates when each transition was triggered. Online resharding
(\sys{}) vs.\ full checkpoint-restart (save to disk, relaunch, reload).}
\label{tab:resharding_overhead}
\begin{tabular}{@{}llrrrr@{}}
\toprule
Config & Transition & Time & Online & Restart & Speedup \\
\midrule
3B   & TP1,PP4,DP2 $\to$ TP1,PP2,DP4   & 3\,min  &  30\,s &  238\,s &  7.8$\times$ \\
3B   & TP1,PP2,DP4 $\to$ TP1,PP1,DP8   & 22\,min &  56\,s &  231\,s &  4.2$\times$ \\
13B  & TP2,PP8,DP1 $\to$ TP1,PP4,DP4   & 20\,min &  42\,s &  404\,s &  9.7$\times$ \\
32B  & TP2,PP16,DP1 $\to$ TP2,PP8,DP2  & 37\,min &  52\,s &  809\,s & 15.7$\times$ \\
7B   & TP4,PP4,DP2 $\to$ TP1,PP8,DP4   & 15\,min &  49\,s &  112\,s &  2.3$\times$ \\
\bottomrule
\end{tabular}
\end{table}

\subsection{GNS Validation}
\label{subsec:gns_validation}

The gradient noise scale is inherently a noisy estimate because it is estimated
from a single mini-batch and fluctuates substantially, especially
early in training when gradients are large and unstable. Our two-phase
EMA smoothing ($\alpha = 0.95$ then $0.99$, \S\ref{subsec:setup})
reduces short-term noise, but a systematic gap between smoothed GNS
and the true critical batch size remains. Prior work has shown that
GNS can underestimate CBS in some settings~\cite{cbs_revisited}; our
empirically determined calibration factor $c = 2.0$ only partially corrects for this gap.

We also tested pre-conditioned GNS
(PGNS)~\cite{gns,adaptdl}, which replaces the raw gradient covariance
with one pre-conditioned by Adam's optimizer state. In our
runs, PGNS shifted standard GNS by another
multiplicative factor and did not improve decision quality,
consistent with both estimators tracking the same signal at
different scales.

A natural question is whether calibration alone is sufficient, and
whether \sys{}'s Goodput formulation helps beyond a perfectly
calibrated CBS method. If throughput is flat across the relevant batch
range, Goodput reduces to statistical efficiency and \sys{} matches
CBS. In practice, however, \autoref{fig:throughput_coupling} shows
that throughput varies substantially with $\Bg$, especially across
parallelism boundaries. In this regime, CBS maximizes per-sample
efficiency without modeling hardware effects, whereas \sys{} accounts
for both and chooses configurations better adapted to the full
$(S, \Bg, \Bm)$ landscape.

\section{Related Work}
\label{sec:related}

\sys{} lies at the intersection of adaptive batch sizing, goodput-aware
scheduling, automated parallelism, and online statistical efficiency estimation.

\noindent \textbf{Adaptive batch sizing.}
The GNS framework~\cite{gns} motivated online batch-size adaptation based
on the gradient noise-to-signal ratio. Follow-up estimators and heuristics
include AdaScale~\cite{adascale}, CABS~\cite{cabs},
SimiGrad~\cite{simigrad}, AdaBatch~\cite{adabatch},
AdaBatchGrad~\cite{adabatchgrad}, and AdAdaGrad~\cite{adadagrad};
large-batch studies further characterized optimization under changing batch
sizes~\cite{scaling_laws,linear_scaling,dont_decay}. Branching-based
methods estimate CBS more directly~\cite{cbs,cbs_revisited}. All optimize
$\Bg$ for statistical efficiency without modeling the 3D-parallelism
throughput surface.

\noindent \textbf{Goodput-aware scheduling.}
Pollux~\cite{adaptdl} introduced goodput as the product of statistical
efficiency and throughput, co-adapting batch size and resource allocation in
a shared DP-only cluster. \sys{} targets a fixed-resource setting where the
decision is the full $(S,\Bg,\Bm)$ tuple under 3D parallelism, with an
LR-aware goodput formulation for Adam-based LLM training.

\noindent \textbf{Automated parallelism and runtime reconfiguration.}
Megatron-LM~\cite{megatron-lm,large_megatron-lm},
DeepSpeed~\cite{deepspeed,zero}, and PipeDream~\cite{pipedream}
provide widely used mechanisms for model, pipeline, and data
parallelism in large-scale training. Automated parallelism planners such
as GSPMD~\cite{gspmd}, Alpa~\cite{alpa}, FlexFlow~\cite{flexflow},
Unity~\cite{unity}, Galvatron~\cite{galvatron}, Merak~\cite{merak},
and nnScaler~\cite{nnscaler} optimize parallel execution strategies for
fixed training configurations. Rubick~\cite{rubick} also exploits job
reconfigurability, but as a cluster scheduler: it co-optimizes execution
plans and multi-resource allocations across jobs, whereas \sys{} changes
$(S,\Bg,\Bm)$ within a single fixed-resource training run as the
preferred batch regime evolves. Elastic systems (Gandiva~\cite{gandiva},
AntMan~\cite{antman}, EasyScale~\cite{easyscale}) reshard in response to
resource changes, not optimization dynamics. HotSPa~\cite{hotspa} switches
strategies within a step, transferring 16-bit parameters and gradients
while keeping the 32-bit optimizer state in one layout; \sys{} switches
between steps and persistently reshards the full state including the
optimizer. RLHF systems such as HybridFlow~\cite{hybridflow} also
perform resharding between training and generation phases, where the
same actor model alternates between workloads with different parallelism
needs. This use case is complementary to \sys{}: HybridFlow reshards
across stages of an RLHF dataflow, whereas \sys{} changes the active
parallelism strategy within a single pretraining run in response to the
Goodput objective. Universal Checkpointing~\cite{ucp} and
ByteCheckpoint~\cite{bytecheckpoint} enable cross-topology resume but do
not decide when to change $(S,\Bg,\Bm)$. \sys{} treats changes in the
preferred batch regime as the trigger for parallelism changes.

\noindent \textbf{GNS estimation.}
Per-example norms~\cite{per_example_gns} and LayerNorm
proxies~\cite{layernorm} improve GNS estimation fidelity or cost but do not address
how to combine the signal with hardware throughput under 3D parallelism.
\sys{} uses GNS as input to a goodput optimizer rather than as a
standalone batch-size selector.

\section{Discussion and Limitations}
\label{sec:discussion}

\paragraph{Decision space.}
\sys{} currently optimizes over data, tensor, and pipeline parallelism,
the global batch size, and the micro-batch size. Other dimensions
commonly used in LLM training, including ZeRO-style optimizer sharding,
context parallelism, sequence parallelism, and activation
checkpointing, are not part of the decision space. Adding these would
enlarge the throughput table and the candidate set but would not change
the Goodput objective itself; the principle of maximizing throughput
times statistical efficiency applies regardless of which knobs are
tuned.

\paragraph{GNS calibration.}
The linear calibration factor $c$ that relates the raw GNS to the
true critical batch size is an approximation. The real relationship
may be model-dependent, training-stage-dependent, or nonlinear.
We found $c = 2.0$ to work well across our four configurations, but
automatically determining $c$, or replacing it with a more principled
estimator, is future work. Furthermore, GNS is intrinsically noisy; our
EMA smoothing and switching margin mitigate noise-driven decisions but
do not eliminate them entirely.

\paragraph{Throughput profiling.}
Our system depends on an offline throughput profile generated once per
model-hardware pair. Prior work on online throughput modeling~\cite{galvatron,adaptdl}
could remove this step and make the system fully self-contained. In
practice, the profiled results can be reused across runs on the same
cluster and model, so this one-time cost is amortized.

\paragraph{Scale and metrics.}
We evaluate on 1--4 nodes (8--32 GPUs) with models up to 32B parameters.
At larger scales, the coupling may be stronger, and benefits may be
larger, but we could not verify this due to compute limits. We report
training loss rather than downstream accuracy because our token budgets
target the adaptive regime rather than full pre-training. Recent work
also suggests that no single LR scaling rule works across all batch
regimes~\cite{surge_lr,timetransfer}, so automatic learning-rate selection
per batch size is another direction for future work.

\paragraph{Short development runs.}
Modern LLM development includes many short proxy pre-training runs before the full-scale run. Practitioners train small models to select data mixtures~\cite{doremi,regmix}, predict data-quality decisions at larger scale~\cite{datadecide,evans2024datacurationjointexample}, or design token-level curricula~\cite{irreducible_curriculum}. These exploratory workloads spend most or all of their lifetime in the earliest, smallest-batch regime of training, the phase where \sys{} already shows the largest gains. Co-adaptive batch-size and parallelism tuning can therefore benefit each proxy run, and the savings compound across the full search process.

\section{Conclusion}
\label{sec:conclusion}

This paper addresses the observation that in 3D-parallel LLM training,
the global batch size and the parallelism strategy are
interdependent: the throughput-optimal parallelism shifts as the batch
size evolves, so any method that fixes one while adapting the other
leaves performance on the table.

We presented \sys{}, the first system to co-adapt the global batch
size, micro-batch size, and 3D parallelism strategy during training,
guided by a Goodput objective that jointly accounts for hardware
throughput and statistical efficiency. \sys{} combines online GNS
estimation under 3D parallelism with throughput-aware candidate
evaluation to continuously select the best configuration, and
supports in-process parallelism reconfiguration with full optimizer
state resharding.

Across four configurations spanning 3B to 32B parameters on both
NVIDIA H100 and AMD MI210 hardware, \sys{} achieves average
time-to-convergence speedups of 3.9--8.0\% over the fastest
individual baseline at each loss threshold (including system
overheads), with peak gains of 11.1\%. Eliminating resharding overhead
entirely would raise these to 4.7--11.4\% average and 13.2\% peak,
indicating clear headroom for further systems engineering.
The analysis confirms that these gains arise from
keeping both throughput and statistical efficiency simultaneously
high, a property that no fixed-parallelism baseline achieves.

Looking ahead, extending the decision space to additional parallelism
dimensions (expert, context, sequence parallelism) and replacing the
offline throughput profile with online modeling are natural next steps
toward fully autonomous training configuration.

\bibliographystyle{ACM-Reference-Format}
\bibliography{references}

\clearpage
\appendix
\begin{center}
  {\LARGE\bfseries Appendix}
\end{center}
\vspace{1em}
\section{Implementation Details}
\label{appendix:implementation}

\subsection{Training Core Modifications}
\label{appendix:impl_training_core}

The training process extends Megatron-LM in three places. First,
we add a GNS manager to the forward--backward loop. Backward
hooks capture the per-microbatch statistics required by the
3D-parallel-aware estimator;
the resulting signal and noise statistics are reduced to rank~0,
smoothed, and sent to the orchestrator.

Second, we replace Megatron's static batching path with an
adaptive one that changes $\Bg$ and $\Bm$ between optimizer
steps without restarting the job or resetting dataset order.
When the orchestrator issues Scale-BS, rank~0 broadcasts the
target $(\Bg,\Bm)$, each rank rebuilds its microbatch
calculator, updates gradient-accumulation state, and rescales
the learning rate. This is
the common case and adds negligible overhead.

Third, we separate the control plane from the computation path.
Only rank~0 communicates with the orchestrator; once a new
configuration is selected, rank~0 broadcasts the command and
parameters so that all workers transition atomically at the same
optimizer-step boundary.

\subsection{Online Parallel Reconfiguration}
\label{appendix:impl_reconfiguration}

Changing the parallelism strategy is the more involved
reconfiguration path. Because each $(d,t,p)$ configuration
determines the shard layout of persistent training state, a
strategy switch must reshard both model weights and optimizer
state before training can resume under the new topology.

We perform this as an in-process operation between optimizer
steps, using host memory as a transient staging layer. The
pipeline extracts source shards, stages them in CPU memory,
tears down and reconstructs NCCL process groups for the target
topology, rebuilds model and optimizer under the new groups,
and loads the staged state into the target shard layout. This
avoids disk I/O and keeps memory usage within that of a single
active configuration. We use Megatron's
\texttt{sharded\_state\_dict()} interface to compute
source--target shard overlaps and derive the required
point-to-point transfers. The full five-phase pipeline is
detailed in Appendix~\ref{appendix:resharding}.

\subsection{Throughput Lookup Table}
\label{appendix:impl_throughput_table}

Our throughput table is generated offline by enumerating valid
$(S,\Bg,\Bm)$ configurations for a fixed model--hardware pair,
subject to divisibility and memory constraints, and running a
short benchmark for each candidate. OOM configurations are
pruned. The resulting lookup table is indexed by $(S,\Bg,\Bm)$
and reused across runs on the same model and cluster.

\subsection{Orchestrator}
\label{appendix:impl_orchestrator}

The orchestrator runs as an out-of-process Python service. On
the training side, rank~0 maintains a non-blocking queue for
sending smoothed GNS measurements and receiving WebSocket
commands. The service periodically evaluates candidates using
the LR-aware Goodput objective, applies the switching margin,
and returns one of three actions: No-Op, Scale-BS, or
Reconfigure.

A Reconfigure command contains the target parallelism
configuration and is executed collectively by all ranks at the
next optimizer-step boundary. The orchestrator also records the
observed latency of recent online reconfigurations and uses it
as $c_{\mathrm{reconfig}}$ in the cost-benefit test,
keeping the switching policy
tied to measured runtime cost rather than a fixed constant.

\section{Online Reconfiguration Pipeline}
\label{appendix:resharding}

We expand on the online reconfiguration mechanism from
\S\ref{subsec:reconfiguration}, focusing on how \sys{} preserves
training state while replacing the active process-group topology.

\begin{figure}[t]
    \centering
    \includegraphics[width=\columnwidth]{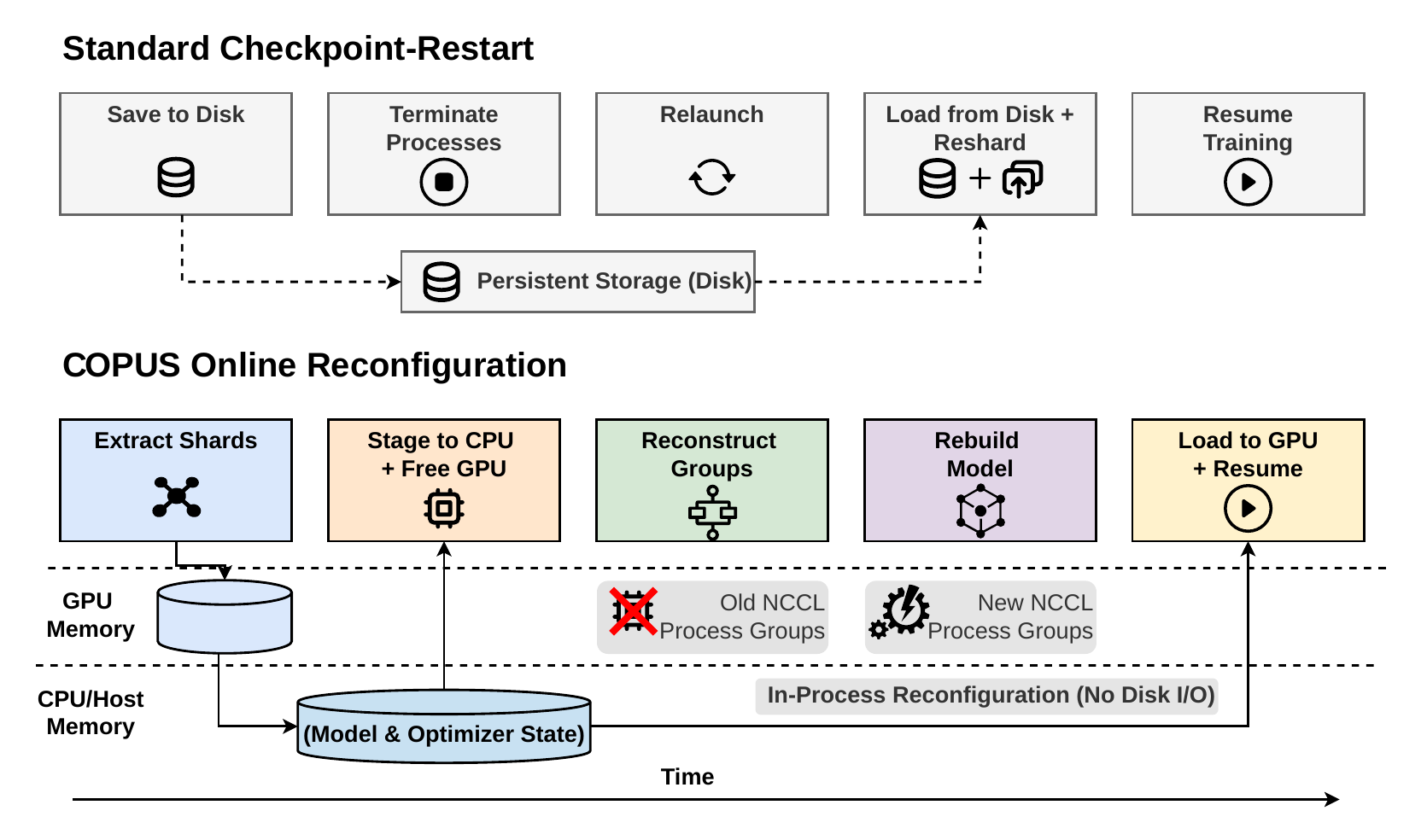}
    \caption{Checkpoint-restart (top) vs.\ \sys{} online reconfiguration (bottom). The standard approach writes to disk, terminates, relaunches, and reloads. \sys{} stages state in host memory and reconstructs process groups in-process, avoiding disk I/O.}
    \label{fig:resharding_pipeline}
\end{figure}

Megatron-LM encodes parallelism through stateful process groups that determine collective communication, optimizer sharding, and buffer layout. Two different 3D-parallel configurations therefore cannot remain simultaneously active inside the same runtime state. Our online reconfiguration pipeline consists of five phases (\autoref{fig:resharding_pipeline}):

\begin{enumerate}
  \item Extract the source shards of the current model weights
  and optimizer state;
  \item Stage those shards in host memory and release the
  current model/optimizer GPU state;
  \item Destroy the old process groups and reconstruct the
  groups for the target topology;
  \item Rebuild the model and optimizer under the new groups;
  \item Load the staged state into the target shard layout and
  resume training.
\end{enumerate}

Host-memory staging is necessary because the source shard data must survive the process-group reconstruction gap, but keeping both the old and new layouts resident on GPU would exceed the memory budget of a single active configuration. Using CPU memory as the transient staging layer lets us keep the reconfiguration online without writing checkpoint files or relaunching the job.

To map tensors from the source layout to the target layout, we use Megatron's \texttt{sharded\_state\_dict()} interface. Each rank materializes metadata for its local shards, including the tensor key, global shape, global offset, and local shape. The reconfiguration planner gathers this metadata across ranks and computes source--target shard overlaps to derive the required point-to-point transfers. This lets the system reconstruct the target shard layout directly from the distributed state exposed by Megatron-LM, without assuming an external algebraic description of tensor partitioning.

\section{Goodput vs.\ CBS Under Scaling Factor Uncertainty}
\label{appendix:goodput_vs_cbs}

Suppose the true critical batch size is $\Bcrit$, but the
measured GNS statistics imply a miscalibrated value
$\widetilde{\Bcrit}=c\,\Bcrit$ for some unknown factor $c$.
A CBS-style rule then selects
\begin{equation}
  {\Bg^\star}_{\mathrm{CBS}} = c\,\Bcrit,
\end{equation}
so the batch-size error scales linearly with the calibration
error.

Now consider a standard saturating throughput model
\begin{equation}
  \mathrm{T}(\Bg)
  = T_{\max}\frac{\Bg}{\Bg + B_{\mathrm{hw}}},
\end{equation}
where $B_{\mathrm{hw}}$ is the batch size at which hardware
throughput begins to saturate. Combining this with
the SE formula $\SE(\Bg) = (1+\GNS)/(\Bg+\GNS)$, the Goodput objective under the same
miscalibrated estimate is, up to constants independent of
$\Bg$,
\begin{equation}
  \mathrm{Goodput}(\Bg)
  \propto
  \frac{\Bg}{(\Bg + B_{\mathrm{hw}})(\Bg + c\Bcrit)}.
\end{equation}
Maximizing yields
\begin{equation}
  {\Bg^\star}_{\mathrm{Goodput}}
  = \sqrt{B_{\mathrm{hw}}\cdot c\Bcrit}.
\end{equation}
The selected batch size is the geometric mean of a
hardware constraint and a statistical one. Its elasticity to the
calibration factor is $1/2$: a factor-$k$ error in $c$ produces
only a factor-$\sqrt{k}$ error in the selected batch size,
rather than the factor-$k$ error of CBS.

This analysis is intentionally simplified. Real throughput under
3D parallelism is discrete rather than smooth and can exhibit
jumps when the throughput-optimal strategy changes across
$(S,\Bg,\Bm)$ configurations. The exact square-root
error reduction therefore does not hold universally, but the same
intuition still applies.

\section{Throughput Profiling Details}
\label{appendix:throughput_profiling}

To evaluate Goodput, the orchestrator needs the throughput $\mathrm{T}(S, \Bg, \Bm, H)$ for every candidate configuration. Throughput depends on the parallelism strategy, batch decomposition, and hardware in ways that are hard to model analytically. Pipeline bubbles, collective communication costs, and memory pressure interact differently across configurations. We measure throughput empirically instead.

Before training, we enumerate all valid $(S, \Bg, \Bm)$ configurations where $d \times t \times p = N_{\text{GPUs}}$ and the model fits in GPU memory. We run a short benchmark (a few training iterations) for each and record the measured throughput. Configurations that run out of memory are pruned. The result is a lookup table indexed by $(S, \Bg, \Bm)$ that the orchestrator queries at each decision point. We build this table once per model-hardware pair and reuse it across training runs.

Rather than building an analytical or simulator-based cost model as in
prior parallelism planners~\cite{alpa,galvatron,flexflow}, we directly benchmark
feasible candidates. This is simpler and captures hardware effects, at
the cost of an offline profiling pass per model-hardware pair. The
profiling step itself is not a contribution of \sys{}; what is new is
feeding this table into the Goodput optimizer so that throughput enters
the batch size decision continuously, not just as a one-time static
parallelism selection.

\section{Loss Spikes}
\label{appendix:loss_spikes}

The loss curves show occasional spikes,
most visibly at batch size transitions but also during steady-state
training. The transition spikes have a direct cause. When the batch
size increases, the learning rate scales up, but Adam's
squared-gradient running average decays at rate $\beta_2$ and still
reflects the previous gradient scale. This pushes the preconditioned
step size past the stability threshold for a few steps until the
second-moment estimate catches up~\cite{adam_loss_spikes}.

Spikes unrelated to batch size changes are common in large-scale
pre-training. During PaLM 540B training, Chowdhery et
al.~\cite{palm} observed roughly 20 spikes and recovered by rolling
back to a checkpoint about 100 steps before the spike and skipping
the triggering data batch. The OPT training logs~\cite{opt} report
similar instabilities, handled by lowering the learning rate before
restarting. More recent work distinguishes narrow spikes that recover
on their own from wide spikes that require intervention, and only
triggers automatic rollbacks for the
latter~\cite{k2v2,spike_no_more}. All spikes we observe are narrow
and recover within a few steps without intervention.

\section{Additional Evaluation Figures}
\label{appendix:additional_trajectories}

\begin{figure*}[p]
    \centering
    {\small\textbf{13B / 2$\times$8 H100}}\par\vspace{0.3em}
    \includegraphics[width=0.62\textwidth]{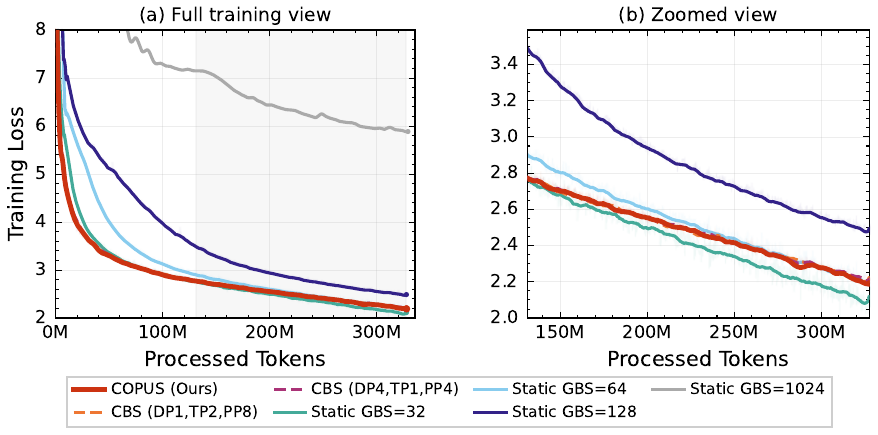}
    \vspace{0.3em}

    {\small\textbf{32B / 4$\times$8 H100}}\par\vspace{0.3em}
    \includegraphics[width=0.62\textwidth]{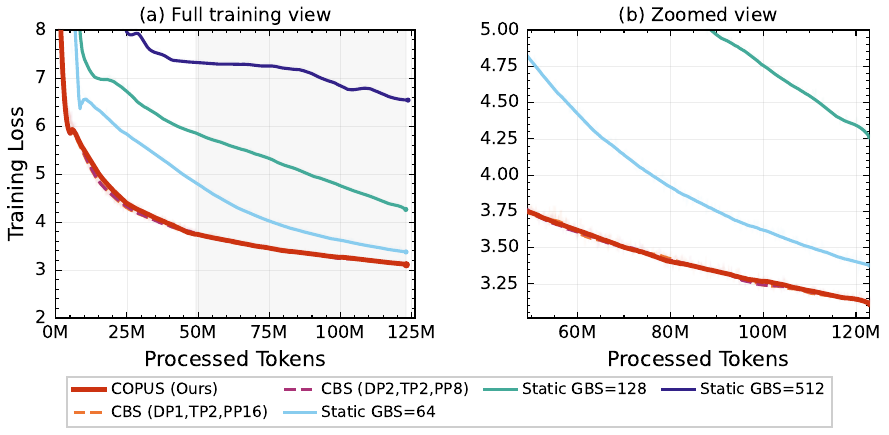}
    \vspace{0.3em}

    {\small\textbf{7B / 4$\times$8 MI210}}\par\vspace{0.3em}
    \includegraphics[width=0.62\textwidth]{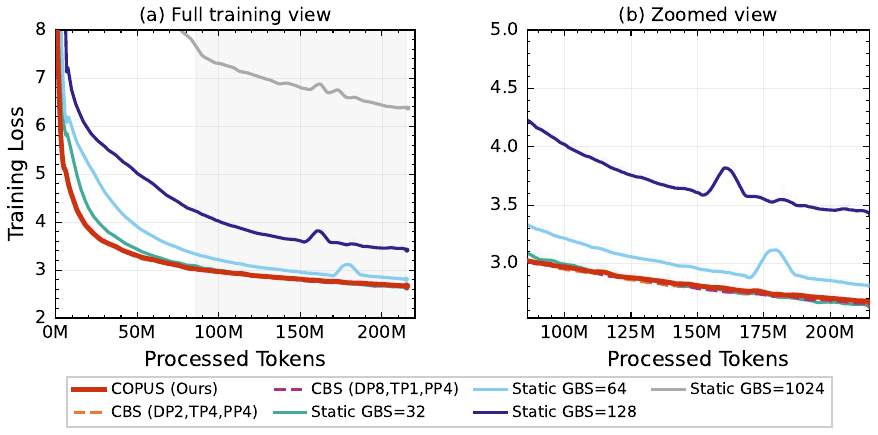}
    \caption{Training loss vs.\ processed tokens for the remaining configurations. Each plot is labeled by model and hardware configuration, and follows the same format as \autoref{fig:loss_vs_tokens}: samples are converted to tokens using a sequence length of 2{,}048, isolating statistical efficiency from throughput.}
    \label{fig:additional_loss_vs_tokens}
\end{figure*}

\autoref{fig:additional_loss_vs_tokens} shows the loss-versus-token
view for the three configurations not shown in
\autoref{fig:loss_vs_tokens}.

\begin{figure*}[t]
    \centering
    {\small\textbf{13B / 2$\times$8 H100}}\par\vspace{0.3em}
    \includegraphics[width=0.48\textwidth]{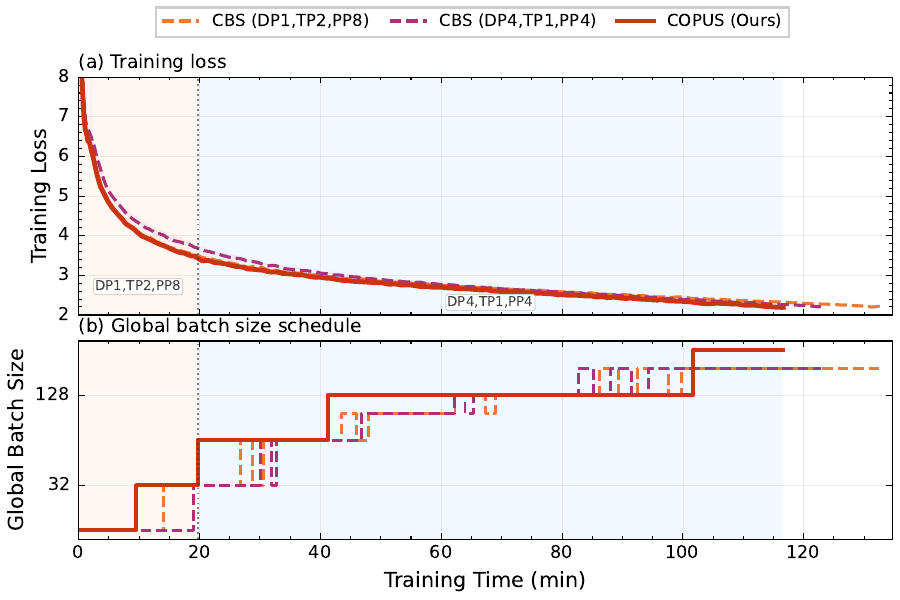}
    \vspace{0.3em}

    {\small\textbf{32B / 4$\times$8 H100}}\par\vspace{0.3em}
    \includegraphics[width=0.48\textwidth]{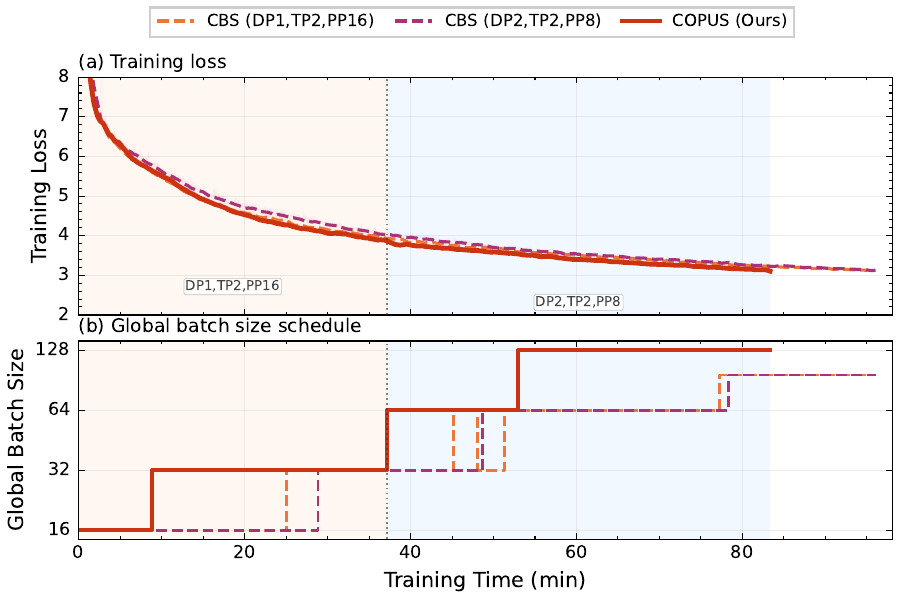}
    \vspace{0.3em}

    {\small\textbf{7B / 4$\times$8 MI210}}\par\vspace{0.3em}
    \includegraphics[width=0.48\textwidth]{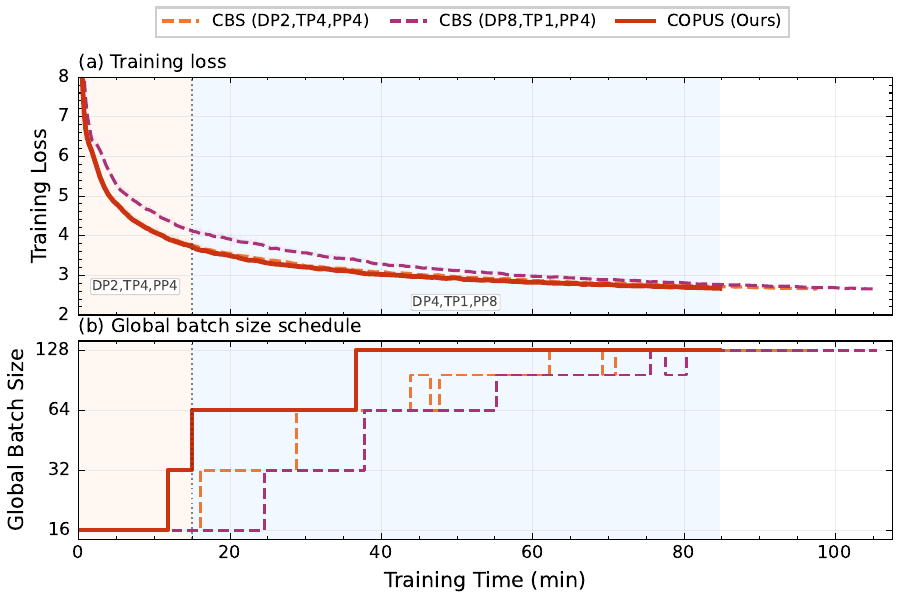}
    \caption{\sys{} adaptation trajectories for the remaining configurations. Each plot is labeled by model and hardware configuration, and follows the same format as \autoref{fig:behavior_trajectory}: loss, batch size schedule, and selected parallelism strategy over training time.}
    \label{fig:additional_behavior_trajectories}
\end{figure*}

\autoref{fig:additional_behavior_trajectories} shows the same
adaptation-trajectory view for the three configurations not shown in
\autoref{fig:behavior_trajectory}.

\begin{figure*}[p]
    \centering
    {\small\textbf{13B / 2$\times$8 H100}}\par\vspace{0.3em}
    \includegraphics[width=0.35\textwidth]{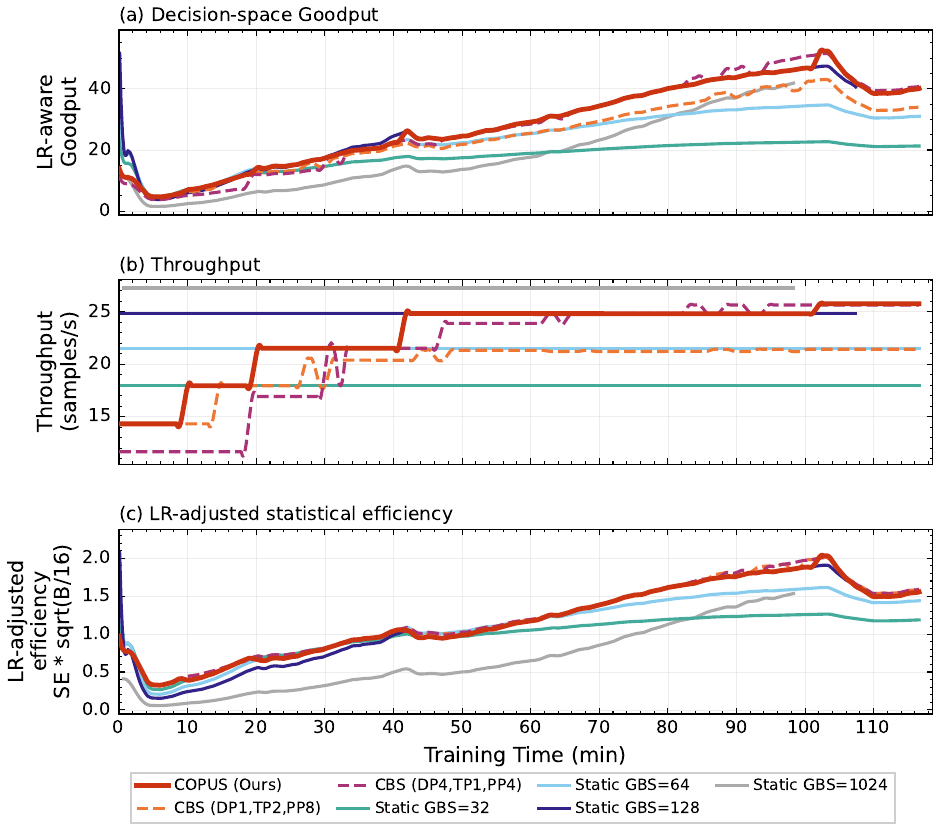}
    \vspace{0.3em}

    {\small\textbf{32B / 4$\times$8 H100}}\par\vspace{0.3em}
    \includegraphics[width=0.35\textwidth]{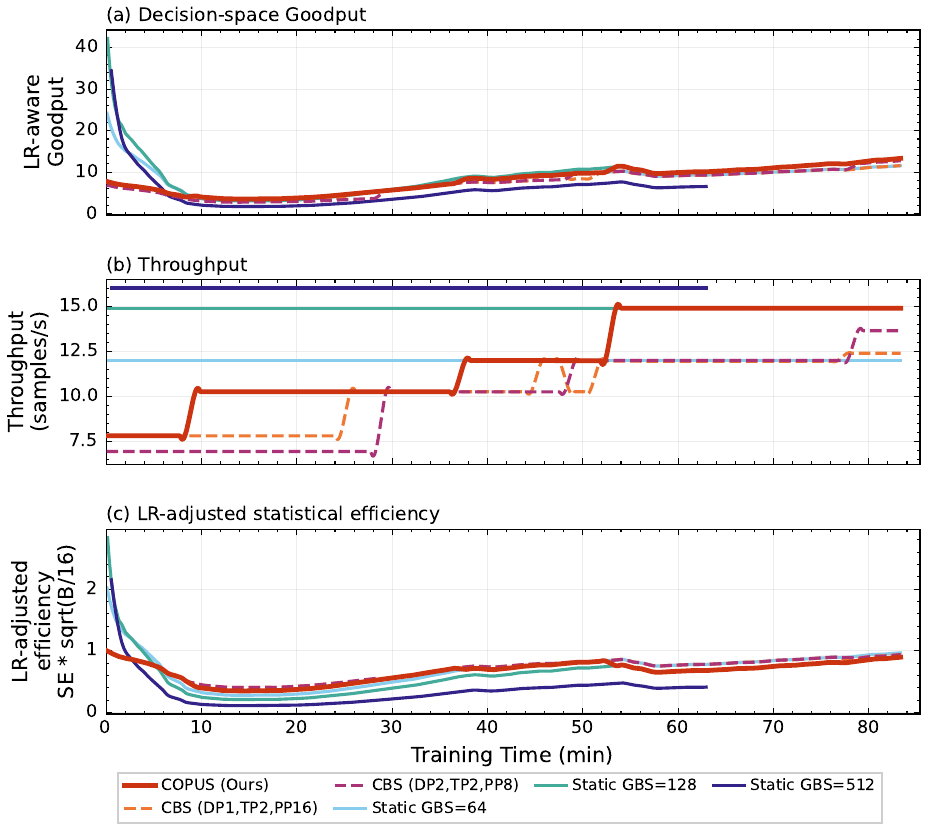}
    \vspace{0.3em}

    {\small\textbf{7B / 4$\times$8 MI210}}\par\vspace{0.3em}
    \includegraphics[width=0.35\textwidth]{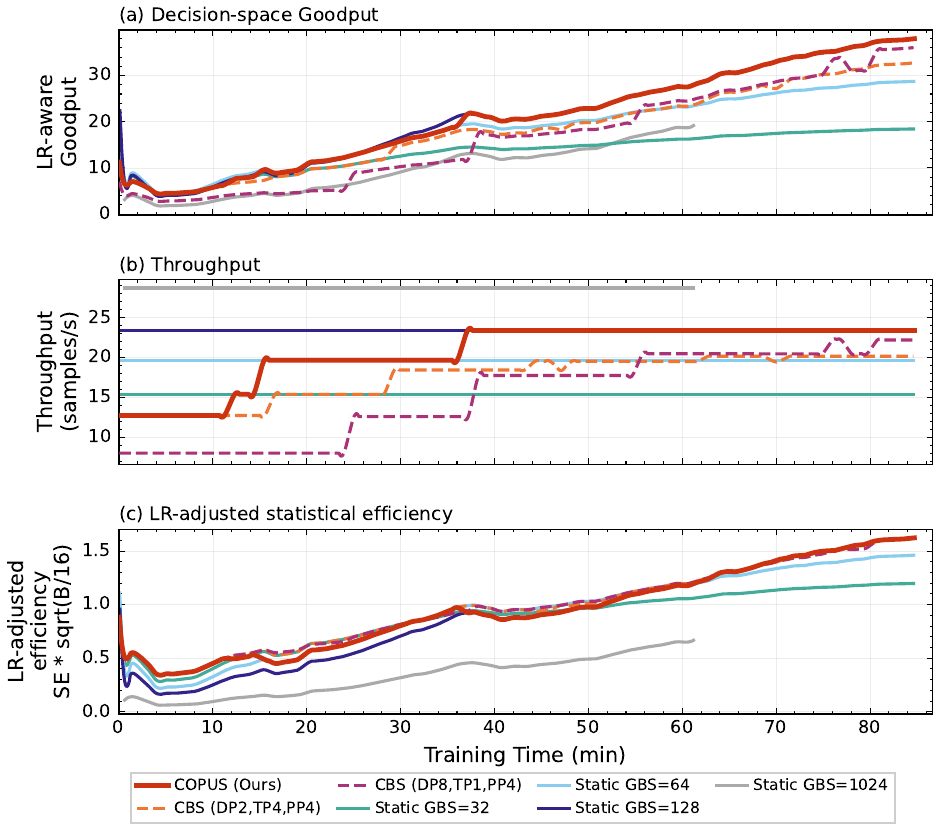}
    \caption{Decision-space Goodput decompositions for the remaining configurations. Each plot is labeled by model and hardware configuration, and evaluates every policy against the \sys{}-observed GNS trajectory for that configuration, using the LR-aware Goodput objective from \autoref{eq:goodput_lr}. The x-axis is limited to the interval where the corresponding \sys{} GNS trajectory is available.}
    \label{fig:additional_goodput_decompositions}
\end{figure*}

\autoref{fig:additional_goodput_decompositions} shows the
corresponding decision-space Goodput decompositions for these
configurations, following the same construction as
\autoref{fig:goodput_vs_time}.

\end{document}